\documentclass[preprint,authoryear,3p]{elsarticle}

\newcommand{\ljump}{\llbracket}
\newcommand{\rjump}{\rrbracket}

\newcommand{\jump}[1]{\ljump #1 \rjump}
\newcommand{\av}[1]{\langle #1 \rangle}

\newcommand{\mH}{\mathcal{H}_{33}}

\usepackage{amsmath}
\usepackage{amssymb}
\usepackage{stmaryrd}
\usepackage{epstopdf}
\usepackage{natbib}
\usepackage{pstricks}
\usepackage{pst-node}
\usepackage{graphicx}
\usepackage{tikz}
\usetikzlibrary{arrows}
\usetikzlibrary{decorations}

\definecolor{corr}{RGB}{255,1,1}

\journal{International Journal of Solids and Structures}

\begin{document}

\begin{frontmatter}



\title{Integral identities for an interfacial crack in an anisotropic bimaterial with an imperfect interface}


\author[aber]{L. Pryce}
\author[aber]{A. Vellender\corref{cor1}}
\author[aber]{A. Zagnetko}

\address[aber]{Department of Mathematics, Aberystwyth University, Physical Sciences Building, Aberystwyth, Ceredigion, Wales, SY23 3BZ}
 \cortext[cor1]{Corresponding author. {\em Tel}: +44 (0)1970622776. {\em Email address}: asv2@aber.ac.uk (A. Vellender)}

\begin{abstract}
We study a crack lying along an imperfect interface in an anisotropic bimaterial. A method is devised where known weight functions for the perfect interface problem are used to obtain singular integral equations relating the tractions and displacements for both the in-plane and out-of-plane fields. The integral equations for the out-of-plane problem are solved numerically for orthotropic bimaterials with differing orientations of anisotropy and for different extents of interfacial imperfection. These results are then compared with finite element computations.
\end{abstract}


\end{frontmatter}



\section{Introduction}
Singular integral equations have played a significant role in the study of crack propagation in elastic media since their introduction by \citet{Mus} and have garnered much scientific attention \citep{Sneddon}. They have been used in the analysis of crack problems in complex domains containing an arbitrary number of wedges and layers separated by imperfect interfaces \citep{Mishuris1997a,Mishuris1997b}; the resulting singular integral equations with fixed point singularities have been analysed by \citet{Duduchava}, based on the theory of linear singular operators \citep{GohbergKrein}. 
More recently, singular integral equations have been applied to problems involving interfacial cracks in both isotropic \citep{Piccolroaz13,AAG4} and anisotropic bimaterials \citep{YuSuo,Morini2}. This paper extends the singular integral equation approach to an anisotropic bimaterial containing an imperfect interface.

Interfacial problems concerning a semi-infinite crack along a perfect interface in an anisotropic bimaterial have been considered in \citet{Suo} through the use of the formalisms proposed by \citet{Stroh} and \citet{Lek}. Expressions were found for the stress intensity factors at the crack tip under the restriction of symmetric loading on the crack faces. Using weight function techniques introduced by \citet{Bueckner1} and developed further by \citet{WillisMovchan}, an approach was developed to find stress intensity factors for an interfacial crack along a perfect interface under asymmetric loading for both the static and dynamic cases, see \citet{Morini} and \citet{Pryce1} respectively.
More widely, weight functions are well developed in the literature for a wide range of fractured body geometries and allow for the evaluation of important constants that may act as fracture criteria. For instance, weight functions have been obtained for a corner crack in a plate of finite thickness \citep{ZhengGlinka}, a 3D semi-infinite crack in an infinite body \citep{KassirSih} and a crack lying perpendicular to the interface in a thin surface layer \citep{Fett}.

Imperfect interfaces provide a more physically realistic interpretation of a bimaterial than a perfect one, accounting for the fact that the interface between two materials is rarely sharp. \citet{Atkinson} took this into account by suggesting the interface be replaced with a thin strip of finite thickness, which provided the bonding material occupying the strip is sufficiently soft may be replaced by so-called imperfect interface transmission conditions. These allow for an interfacial displacement jump in direct proportion to the traction, which is itself continuous across the interface \citep{Antipov2001,Lenci,Mishuris2001}. Such transmission conditions alter physical fields near the crack tip significantly; for instance the usual perfect interface square root stress singularity is no longer present and is instead replaced by a logarithmic singularity \citep{MisKuhn}, although tractions remain bounded along the interface. More general imperfect interface transmission conditions were derived by \citet{
BenvenisteMiloh} which considered a thin curved isotropic layer of constant thickness, while \citet{Benveniste} presented a general interface model for a 3D arbitrarily curved thin anisotropic interphase between two anisotropic solids. 

Weight function techniques have \textcolor{black}{been recently} adapted to imperfect interface settings to 
quantify crack tip asymptotics in thin domains \citep{Vellender1}, analyse problems of waves in thin waveguides \citep{Vellender2} and conduct perturbation analysis for large imperfectly bound bimaterials containing small defects \textcolor{black}{\citep{Vellender3}}; the absence of the square root singularity means that the weight functions are not used to find stress intensity factors, but instead yield asymptotic constants which describe the crack tip opening displacement. This quantity was proposed for use in fracture criteria by \citet{Wells} and \citet{Cottrell} and later justified rigorously by \citet{RiceSorenson}, \citet{Shih} and \citet{Kanninen}. Despite their great utility, the derivation of such weight functions is often not straightforward and so the approach deployed in the remainder of this paper efficiently utilises existing relationships between known weight functions without the need to derive further expressions.

\textcolor{black}{The problem considered here is the anisotropic equivalent of that seen in \citet{AAG4}, which considered solely isotropic bimaterials. Besides this, perhaps the key novel feature in the present manuscript from a methodology viewpoint, is that known weight functions derived for the {\em perfect} interface problem are used in the derivation of singular integral equations for the {\em soft imperfect} interface case. This differs from previous approaches; for instance \citet{AAG4} used specially-derived weight functions that took into account the local crack-tip behaviour brought about by the presence of imperfect interface transmission conditions, whereas the approach employed here uses existing perfect interface weight functions, which have fundamentally different behaviour near the crack tip to the physical solution in the imperfect interface problem.}

The paper is structured as follows: Section 2 introduces the problem geometry and model for the imperfect interface. In Section 3, previously found results used in the derivation of the singular integral equations are discussed. These include the weight functions derived using the method of \citet{WillisMovchan} and the Betti formula which can be used to relate the weight functions to the physical fields along both the crack and imperfect interface. Section 4 concentrates on solving the out-of-plane (mode III) problem. Singular integral equations are derived and used to obtain the displacement jump across both the crack and interface for a number of orthotropic bimaterials with varying levels of interface imperfection. Finite element methods for the same physical problems are also used to obtain the same results and then a comparison is made between the results obtained from the two opposing methods. The in-plane problem is considered in Section 5, where singular 
integral equations are obtained for the mode I and mode II tractions and displacements and some computations are performed.

\section{Problem formulation}
We consider an infinite anisotropic bimaterial with an imperfect interface and a semi-infinite interfacial crack respectively lying along the positive and negative $x_1$ semi-axes. The materials above and below the $x_1$-axis will be denoted materials I and II respectively.

\begin{figure}
\begin{center}
\begin{tikzpicture}[scale=2.8]
\draw [-] (0,0) -- (2,0);
\draw [-] (-2,0.1) -- (0,0);
\draw [-] (-2,-0.1) -- (0,0);
\draw [thick, <->] (0, 1.3) -- (0,0) -- (1.3, 0);
\node [left] at (0,1.3) {$x_2$};
\node [below] at (1.3,0) {$x_1$};
\node [above] at (1.5,0.7) {I};
\node [below] at (1.5,-0.7) {II};
\draw [-,red] (-1.5,0.075) to[out=90,in=90] (-0.5,0.025);
\draw [->,red] (-1,0.05) -- (-1,0.35);
\draw [->,red] (-1.25,0.0625) -- (-1.25,0.29);
\draw [->,red] (-0.75,0.0375) -- (-0.75,0.28);
\node [right] at (-0.5,0.2) {$p^+$};
\draw [-,blue] (-1.8,-0.09) to[out=-90,in=-90] (-0.8,-0.04);
\draw [->,blue] (-1.3,-0.065) -- (-1.3,-0.35);
\draw [->,blue] (-1.55,-0.0775) -- (-1.55,-0.29);
\draw [->,blue] (-1.05,-0.0525) -- (-1.05,-0.28);
\node [right] at(-0.8,-0.2) {$p^-$};
\end{tikzpicture}
\caption{Geometry}
\label{geometry}
\end{center}
\end{figure}
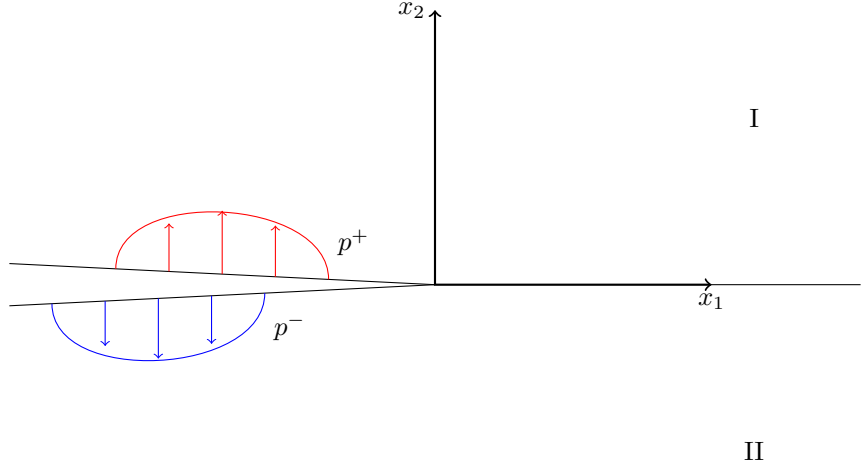

The imperfect interface transmission conditions for $x_1>0$ are given by
\begin{equation}\label{tractrans}
\mathbf{t}(x_1,0^+)=\mathbf{t}(x_1,0^-),
\end{equation}\begin{equation}
\mathbf{u}(x_1,0^+)-\mathbf{u}(x_1,0^-)=\mathbf{K}\mathbf{t}(x_1,0^+),\label{jumptrans}
\end{equation}
where $\mathbf{t}=(t_1,t_2,t_3)^T=(\sigma_{21},\sigma_{22},\sigma_{23})^T$ is the traction vector and $\mathbf{u}=(u_1,u_2,u_3)^T$ is the displacement vector. The matrix $\mathbf{K}$ quantifies the extent of imperfection of the interface, with $\mathbf{K}=\mathbf{0}$ corresponding to the perfect interface. For an anisotropic bonding material, $\mathbf{K}$ has the following structure:
\begin{equation}
\mathbf{K}=\begin{pmatrix} K_{11}&K_{12}&0\\K_{12}&K_{22}&0\\0&0&\kappa\end{pmatrix}.
\end{equation}
The loading on the crack faces is considered known and given by
\begin{equation}
\mathbf{t}(x_1,0^+)=\mathbf{p}^+(x_1),\quad\mathbf{t}(x_1,0^-)=\mathbf{p}^-(x_1),\quad\text{for }x_1<0.
\end{equation}
The geometry considered is illustrated in Figure \ref{geometry}. The only restriction imposed on $\mathbf{p}^\pm$ is that they must be self-balanced; note in particular that this allows for discontinuous and/or asymmetric loadings. The symmetric and skew-symmetric parts of the loading are given by $\av{\mathbf{p}}$ and $\jump{\mathbf{p}}$ respectively, where the notation $\av{f}$ and $\jump{f}$ respectively denote the average and jump of the argument function:
\[
\av{f}(x_1)=\frac{1}{2}(f(x_1,0^+)+f(x_1,0^-)),\quad \jump{f}(x_1)=f(x_1,0^+)-f(x_1,0^-).
\]

\section{Application of existing weight functions}
\subsection{Weight functions}
\textcolor{black}{\citet{Bueckner1} defined weight functions as non-trivial singular solutions of the homogeneous traction-free problem. \citet{WillisMovchan} introduced weight functions in a mirrored domain and related physical quantities with the auxiliary weight functions via use of Betti's identity; this procedure has been recently used to derive singular integral equations for isotropic bimaterials joined by an imperfect interface \citep{AAG4}. The approach employed there required the use of weight functions that had been designed for an imperfect interface setting for isotropic bimaterials.
In the spirit of the efficiency outlined in the introduction, we will in this section introduce a method where integral identities for the physical problem with an imperfect interface are found using existing weight functions formulated in a {\em perfect} interface setting. Such weight functions can be found in the paper of \citet{Morini}. Note that such weight functions play a role only as solutions to auxiliary problems and have no immediate physical interpretation; we refer the reader to \cite{WillisMovchan} for further details.}

The weight function used is the solution of the problem with the crack occupying the positive $x_1$ axis with square-root singular displacement at the crack tip, as given in  \citet{Morini}. The transmission conditions for the weight functions for $x_1<0$ are given as
\begin{equation}
\mathbf{\Sigma}(x_1,0^+)=\mathbf{\Sigma}(x_1,0^-),
\end{equation}\begin{equation}\label{eq:contofU}
\mathbf{U}(x_1,0^+)=\mathbf{U}(x_1,0^-),
\end{equation}
where $\mathbf{U}$ is the singular displacement field and $\mathbf{\Sigma}$ is the corresponding traction field. Note in particular that condition (\ref{eq:contofU}) corresponds to a perfect interface weight function problem in contrast to the imperfect interface problem being physically considered.

It was shown in \citet{Morini} that the following equations hold for the Fourier transforms of the symmetric and skew-symmetric parts of the weight function:
\begin{equation}\label{weightsymm}
\jump{\bar{\mathbf{U}}}^+(\xi)=\frac{1}{|\xi|}(i\mathrm{sign}(\xi)\mathrm{Im}(\mathbf{H})-\mathrm{Re}(\mathbf{H}))\av{\bar{\mathbf{\Sigma}}}^-(\xi);
\end{equation}\begin{equation}\label{weightantisymm}
\av{\bar{\mathbf{U}}}(\xi)=\frac{1}{2|\xi|}(i\mathrm{sign}(\xi)\mathrm{Im}(\mathbf{W})-\mathrm{Re}(\mathbf{W}))\av{\bar{\mathbf{\Sigma}}}^-(\xi),
\end{equation}
where $\mathbf{H}=\mathbf{B}_I+\mathbf{B}^\star_{II}$ and $\mathbf{W}=\mathbf{B}_I-\mathbf{B}^\star_{II}$. Here, $\mathbf{B}_{I}$ and $\mathbf{B}_{II}$ are the surface admittance tensors of materials I and II respectively, superscript $^\star$ denotes complex conjugation and bars denote Fourier transforms with respect to $x_1$ defined as
\begin{equation}
 \bar{f}(\xi)=\mathcal{F}[f](\xi)=\int\limits_{-\infty}^\infty f(x_1)e^{i\xi x_1}\mathrm{d}x_1.
\end{equation}
The matrices $\mathbf{H}$ and $\mathbf{W}$ have the form
\begin{equation}\label{HW}
\mathbf{H}=\begin{pmatrix} H_{11}&-i\beta\sqrt{H_{11}H_{22}}&0\\i\beta\sqrt{H_{11}H_{22}}&H_{22}&0\\0&0&H_{33}\end{pmatrix},\quad
\mathbf{W}=\begin{pmatrix} \delta_1 H_{11}&i\gamma\sqrt{H_{11}H_{22}}&0\\-i\gamma\sqrt{H_{11}H_{22}}&\delta_2 H_{22}&0\\0&0&\delta_3 H_{33}\end{pmatrix}.
\end{equation}
{The entries of these matrices can be expressed in terms of the components of the material compliance tensors, $\mathbf{S}$. Explicit expressions for $\mathbf{H}$ and $\mathbf{W}$ for orthotropic bimaterials are given in the appendix.}

\subsection{Betti formula}
In this section, the Betti formula is extended to the case of general asymmetrical loading applied at the crack surfaces. The Betti formula is used in order to relate the physical solution to the weight function, which is a special singular solution to the homogeneous problem with traction-free crack faces \citep{WillisMovchan, Piccolroaz07}.

Applying the Betti formula to a semi-circular domain in the half-plane $x_2>0$, whose straight boundary is the line $x_2=0^+$, and whose radius $R\to\infty$, the following equation is obtained
\begin{equation}
\label{bettiupper}
\int\limits_{(x_2=0^+)}
\Big\{ \boldsymbol{\mathcal{R}} \mathbf{U}(x_1'-x_1,0^+) \cdot \mathbf{t}(x_1,0^+) - \boldsymbol{\mathcal{R}} \mathbf{\Sigma}(x_1'-x_1,0^+) \cdot \mathbf{u}(x_1,0^+) \Big\} \mathrm{d}x_1 = 0.
\end{equation}
where $\boldsymbol{\mathcal{R}}$ is a rotation matrix given by
\[
\begin{pmatrix} -1&0&0\\0&1&0\\0&0&-1\end{pmatrix}.
\]
Another equation can be derived by applying the Betti formula to a semi-circular domain in the half-plane $x_2<0$ and taking the limit, $R\to\infty$, which after some manipulation in the spirit of \citet{Piccolroaz09} for example, yields 
\begin{equation}
\label{recid3}
\boldsymbol{\mathcal{R}} \jump{\mathbf{U}} * \av{\mathbf{t}}^{(+)} - \boldsymbol{\mathcal{R}} \av{\mathbf{\Sigma}}^{(-)} * \jump{\mathbf{u}} =
- \boldsymbol{\mathcal{R}} \jump{\mathbf{U}} * \av{\mathbf{p}} - \boldsymbol{\mathcal{R}} \av{\mathbf{U}} * \jump{\mathbf{p}},
\end{equation}
where the convolutions are taken with respect to $x_1$, that is 
\[
(f*g)(x_1)=\int_{-\infty}^\infty f(x_1-t)g(t)\mathrm{d}t,
\]
and superscripts $^{(\pm)}$ denote the restriction of the preceding function to the respective semi-$x_1$-axis.
Applying Fourier transforms then gives
\begin{equation}\label{finalbetti}
\bar{\jump{\mathbf{U}}}^T \boldsymbol{\mathcal{R}}\bar{\av{\mathbf{t}}}^+ - (\bar{\av{\mathbf{\Sigma}}}^-)^T\boldsymbol{\mathcal{R}}\bar{\jump{\mathbf{u}}}=-\bar{\jump{\mathbf{U}}}^T\boldsymbol{\mathcal{R}}\bar{\av{\mathbf{p}}} - \bar{\av{\mathbf{U}}}^T\boldsymbol{\mathcal{R}}\bar{\jump{\mathbf{p}}}.
\end{equation}

Note that the exact nature of the weight functions $\mathbf{U}$ and $\mathbf{\Sigma}$ used in this analysis have not been specified at this stage and so identity (\ref{finalbetti}) is valid for a large class of weight functions. In particular, this allows for the use of perfect interface weight functions for the imperfect interface physical setting.
In the case of perfect interface physical solution and weight functions, the corresponding analysis has been done in \citet{Piccolroaz13, Morini}, for isotropic and anisotropic materials respectively, while for imperfect interfaces joining isotropic bodies, details can be found in \citet{AAG4}.

\section{Integral identities for mode III}
\subsection{Derivation of integral identities}
We now seek boundary integral equations relating the mode III interfacial traction and displacement jump over the crack in the anisotropic bimaterial. This will utilise the Betti identity in order to relate the physical solution with the perfect interface weight functions.

Considering only the mode III components of \eqref{finalbetti} the following equation holds:
\begin{equation}
\bar{\jump{U}}(\xi)\overline{\av{t}^{(+)}}(\xi) - \overline{\av{\Sigma}^{(-)}}(\xi)\bar{\jump{u}}(\xi)=-\bar{\jump{U}}(\xi)\bar{\av{p}}(\xi) - \bar{\av{U}}(\xi)\bar{\jump{p}}(\xi),
\end{equation}
where the subscripts have been removed for notational brevity. Splitting $\jump{U}$ into the sum of $\jump{U}^{(\pm)}$ and also separating $\jump{u}$  into the sum of $\jump{u}^{(\pm)}$ gives
\begin{align}
\overline{\jump{U}^{(+)}}(\xi)\overline{\av{t}^{(+)}}(\xi) + \overline{\jump{U}^{(-)}}(\xi)\overline{\av{t}^{(+)}}(\xi) - \overline{\av{\Sigma}^{(-)}}(\xi)\overline{\jump{u}^{(+)}}(\xi) -
\overline{\av{\Sigma}^{(-)}}(\xi)\overline{\jump{u}^{(+)}}(\xi)=\nonumber\\
-\bar{\jump{U}}(\xi)\bar{\av{p}}(\xi) - \bar{\av{U}}(\xi)\bar{\jump{p}}(\xi).\label{nocancel}
\end{align}
Note that if imperfect interface weight functions are used, then the second and third terms of the left hand side of (\ref{nocancel}) immediately due to the transmission conditions \citep{Vellender3}. However, using perfect interface weight functions, this is not true.

Using the transmission conditions, $\overline{\jump{U}^{(-)}}=0$ and $\overline{\jump{u}^{(+)}}=\kappa \bar{\av{t}}$ yields
\begin{equation}\label{m3div}
\overline{\av{t}^{(+)}} - \left(\frac{\overline{\av{\Sigma}^{(-)}}}{\overline{\jump{U}^{(+)}} - \kappa\overline{\av{\Sigma}^{(-)}}}\right) \overline{\jump{u}^{(-)}} = -\left(\frac{\bar{\jump{U}}}{\overline{\jump{U}^{(+)}} - \kappa\overline{\av{\Sigma}^{(-)}}}\right)\bar{\av{p}} - \left(\frac{\bar{\av{U}}}{\overline{\jump{U}^{(+)}} - \kappa\overline{\av{\Sigma}^{(-)}}}\right)\bar{\jump{p}}.
\end{equation}

From equations \eqref{weightsymm} and \eqref{weightantisymm} the following relationships hold for the mode III components of the weight functions:
\begin{equation}
\bar{\jump{U}}=\overline{\jump{U}^{(+)}}(\xi)=-\frac{H_{33}}{|\xi|}\overline{\av{\Sigma}^{(-)}}(\xi);\quad \av{\bar{U}}=-\frac{\delta_3 H_{33}}{2|\xi|}\overline{\av{\Sigma}^{(-)}}(\xi)=\frac{\delta_3}{2}\jump{\bar{U}}(\xi);
\end{equation}
when combined with equation \eqref{m3div} the following relationship is obtained:
\begin{equation}\label{ftsol3}
\overline{\av{t}^{(+)}} - A(\xi)\overline{\jump{u}^{(-)}} = -(1+\kappa A(\xi))\bar{\av{p}}-\frac{\delta_3}{2}(1+\kappa A(\xi))\bar{\jump{p}},
\end{equation}
where
\[
A(\xi)=-\frac{|\xi|}{\kappa|\xi|+\kappa\mathcal{H}_{33}},\qquad \mH=\frac{H_{33}}{\kappa}.
\]

Applying the inverse Fourier transform to equation \eqref{ftsol3} for the two cases, $x_1<0$ and $x_1>0$, the following relationships are obtained:
\begin{equation}\label{invm3less0}
\mathcal{F}^{-1}_{x_1<0}\left[A(\xi)\overline{\jump{u}^{(-)}}\right]=\mathcal{F}^{-1}_{x_1<0}\left[(1+\kappa A(\xi))\bar{\av{p}}\right] + \frac{\delta_3}{2}\mathcal{F}^{-1}_{x_1<0}\left[(1+\kappa A(\xi))\bar{\jump{p}}\right];
\end{equation}\begin{equation}\label{invm3gre0}
\av{t}^{(+)}(x_1) = \mathcal{F}^{-1}_{x_1>0}\left[A(\xi)\overline{\jump{u}^{(-)}} \right] - \mathcal{F}^{-1}_{x_1>0}\left[(1+\kappa A(\xi))\bar{\av{p}}\right] - \frac{\delta_3}{2}\mathcal{F}^{-1}_{x_1>0}\left[(1+\kappa A(\xi))\bar{\jump{p}}\right].
\end{equation}

To calculate these inversions the following relationships are used:
\begin{equation}\label{invFTA}
\mathcal{F}^{-1}\left[A(\xi)\bar{f}(\xi)\right]=\frac{1}{\pi\kappa}\left(S_{\mH}\ast f'\right)(x_1);
\end{equation}\begin{equation}\label{invFT1+A}
\mathcal{F}^{-1}\left[(1+\kappa A(\xi))\bar{f}(\xi)\right]=-\frac{\mH}{\pi}\left(T_{\mH}\ast f\right)(x_1),
\end{equation}
where
\begin{equation}\label{S}
S_{\mH}(x_1)=\mathrm{sign}(x_1)\mathrm{si}(\mH|x_1|)\cos(\mH|x_1|) - \mathrm{sign}(x_1)\mathrm{ci}(\mH|x_1|)\sin(\mH|x_1|),
\end{equation}\begin{equation}\label{T}
T_{\mH}(x_1)=\mathrm{si}(\mH|x_1|)\sin(\mH|x_1|) - \mathrm{ci}(\mH|x_1|)\cos(\mH|x_1|),
\end{equation}
and si and ci are the sine and cosine integral functions respectively, given by
\begin{equation}
\mathrm{si}(x_1)=-\int_{x_1}^\infty \frac{\sin t}{t}\mathrm{d}t,\quad
\mathrm{ci}(x_1)=-\int_{x_1}^\infty \frac{\cos t}{t}\mathrm{d}t.
\end{equation}
These functions have the same properties as their counterparts from the isotropic case considered by \citet{AAG4}, but with different constants. In particular, the function $S_{\mH}(x_1)$ behaves as
\begin{equation}
 S_{\mH}(x_1)=-\frac{\pi}{2}\mathrm{sign}(x_1)+O(|x_1|),\:x_1\to0,\qquad S_{\mH}(x_1)=-\frac{\mathrm{sign}(x_1)}{\mH|x_1|}+O\left(\frac{1}{|x_1|^3}\right),\:x_1\to\pm\infty,
\end{equation}
while $T_{\mH}(x_1)$ has behaviour of the form
\begin{equation}
 T_{\mH} (x_1)=\ln(\mH|x_1|)+O(1),\: x_1\to0,\qquad T_{\mH}(x_1)=-\frac{1}{\mH^2|x_1|^2}+O\left(\frac{1}{|x_1|^3}\right),\:x_1\to\pm\infty.
\end{equation}

We introduce convolution operators $\mathcal{S}_{\mH}$ and $\mathcal{T}_{\mH}$, as well as projection operators $\mathcal{P}_\pm$: 
\begin{equation}\label{curlySTdef}
\mathcal{S}_{\mH}\varphi(x_1)=(S_{\mH}\ast\varphi)(x_1),\qquad
\mathcal{T}_{\mH}\varphi(x_1)=(T_{\mH}\ast\varphi)(x_1),
\end{equation}\begin{equation}
\mathcal{P}_\pm \varphi(x_1)=\begin{cases} \varphi(x_1)\quad \pm x_1\ge 0,\\ 0 \quad \mathrm{otherwise},\end{cases}
\end{equation}
in order to rewrite the identities \eqref{invm3less0} and \eqref{invm3gre0} as
\begin{equation}\label{m3inteqneg}
\frac{1}{\pi\kappa}\mathcal{S}_{\mH}^{(s)}\frac{\partial \jump{u}^{(-)}}{\partial x_1} -\frac{1}{\pi\kappa}\jump{u}^{(-)}(0^-)S_{\mH}(x_1)= -\frac{\mH}{\pi}\mathcal{T}_{\mH}^{(s)}\av{p}(x_1) - \frac{\delta_3\mathcal{H}_{33}}{2\pi}\mathcal{T}_{\mH}^{(s)}\jump{p}(x_1),\quad x_1<0,
\end{equation}
\begin{align}\label{m3inteqpos}
\av{t}^{(+)}(x_1)=\frac{1}{\pi\kappa}\mathcal{S}_{\mH}^{(c)}\frac{\partial \jump{u}^{(-)}}{\partial x_1}-\frac{1}{\pi\kappa}\jump{u}^{(-)}(0^-)S_{\mH}(x_1)&+\frac{\mH}{\pi}\mathcal{T}_{\mH}^{(c)}\av{p}(x_1) \nonumber \\&+ \frac{\delta_3\mathcal{H}_{33}}{2\pi}\mathcal{T}_{\mH}^{(c)}\jump{p}(x_1), \quad x_1>0,
\end{align}
where
\begin{equation}
\mathcal{S}_{\mH}^{(s)}=\mathcal{P}_-\mathcal{S}_{\mH}\mathcal{P}_-,\qquad \mathcal{T}_{\mH}^{(s)}=\mathcal{P}_-\mathcal{T}_{\mH}\mathcal{P}_-,
\end{equation}
are singular operators and
\begin{equation}
\mathcal{S}_{\mH}^{(c)}=\mathcal{P}_+\mathcal{S}_{\mH}\mathcal{P}_-,\qquad \mathcal{T}_{\mH}^{(c)}=\mathcal{P}_+\mathcal{T}_{\mH}\mathcal{P}_-,
\end{equation}
are compact. The second term on the left hand side of \eqref{m3inteqneg} and right hand side of \eqref{m3inteqpos} appear as a result of the discontinuity of the derivative of $\jump{u}^{(-)}$ at $x_1=0$.

\subsection{Alternative integral identities}
 The integral identities \eqref{m3inteqneg} and \eqref{m3inteqpos} can be formulated in alternative ways, which depending upon the specific problem parameters and loadings, can aid the ease with which computations may be performed. 
  Combining equations \eqref{invFTA}, \eqref{invFT1+A} and \eqref{curlySTdef} yields the auxiliary relationship
\begin{equation}\label{STrel}
-\frac{\mH}{\pi}\mathcal{T}_{\mH}\varphi = \mathcal{I}\varphi + \frac{1}{\pi}\mathcal{S}_{\mH}\varphi'.
\end{equation}
Using this relationship, equations \eqref{m3inteqneg} and \eqref{m3inteqpos} can be rewritten as follows:
\begin{align}
-\frac{\mH}{\pi\kappa}\mathcal{T}_{\mH}^{(s)}\jump{u}^{(-)} - \frac{1}{\kappa}\jump{u}^{(-)} = &\frac{1}{\pi}\mathcal{S}_{\mH}^{(s)}\frac{\partial\av{p}}{\partial x_1}-\frac{1}{\pi}\av{p}(0^-)S_{\mH}+\av{p} \nonumber\\&+\frac{\delta_3}{2\pi}\mathcal{S}_{\mH}^{(s)}\frac{\partial\jump{p}}{\partial x_1} - \frac{\delta_3}{2\pi}\jump{p}(0^-)S_{\mH} + \frac{\delta_3}{2}\jump{p},\quad x_1<0;\label{altneg}
\end{align}
\begin{align}
\av{t}^{(+)} = -\frac{\mH}{\pi\kappa}\mathcal{T}_{\mH}^{(c)}\jump{u}^{(-)}  - \frac{1}{\pi}\mathcal{S}_{\mH}^{(c)}\frac{\partial\av{p}}{\partial x_1}+&\frac{1}{\pi}\av{p}(0^-)S_{\mH}+\av{p} \nonumber\\&- \frac{\delta_3}{2\pi}\mathcal{S}_{\mH}^{(c)}\frac{\partial\jump{p}}{\partial x_1} + \frac{\delta_3}{2\pi}\jump{p}(0^-)S_{\mH},\quad x_1>0.\label{altpos}
\end{align}
It is also possible to write these equations using only the operator $\mathcal{T}_{\mH}$: 
\begin{equation}
\label{res3b}
-\frac{\mH}{\pi\kappa} \mathcal{T}_{\mH}^{(s)} \jump{u}^{(-)} - \frac{1}{\kappa} \jump{u}^{(-)} =
-\frac{\mH}{\pi} \mathcal{T}_{\mH}^{(s)} \av{p} - \frac{\delta_3\mathcal{H}_{33}}{2\pi} \mathcal{T}_{\mH}^{(s)} \jump{p}, \quad x_1 < 0;
\end{equation}
\begin{equation}
\label{res4b}
\av{t}^{(+)} = 
-\frac{\mH}{\pi\kappa} \mathcal{T}_{\mH}^{(c)} \jump{u}^{(-)} + \frac{\mH}{\pi} \mathcal{T}_{\mH}^{(c)} \av{p} + \frac{\delta_3\mH}{2\pi} \mathcal{T}_{\mH}^{(c)} \jump{p}, \quad x_1 > 0,
\end{equation}
or solely the operator $\mathcal{S}_{\mH}$:
\begin{align}
\label{res5b}
\frac{1}{\pi\kappa}& \mathcal{S}_{\mH}^{(s)} \frac{\partial \jump{u}^{(-)}}{\partial x_1} - \frac{1}{\pi\kappa}\jump{u}^{(-)}(0^-)S_{\mH} =\nonumber\\ &\frac{1}{\pi}\mathcal{S}_{\mH}^{(s)}\frac{\partial\av{p}}{\partial x_1}-\frac{1}{\pi}\av{p}(0^-)S_{\mH}+\av{p} +\frac{\delta_3}{2\pi}\mathcal{S}_{\mH}^{(s)}\frac{\partial\jump{p}}{\partial x_1} - \frac{\delta_3}{2\pi}\jump{p}(0^-)S_{\mH} + \frac{\delta_3}{2}\jump{p}, \quad x_1 < 0;
\end{align}
\begin{align}
\label{res6b}
\av{t}^{(+)} = 
\frac{1}{\pi\kappa} &\mathcal{S}_{\mH}^{(c)} \frac{\partial \jump{u}^{(-)}}{\partial x_1} - \frac{1}{\pi\kappa}\jump{u}^{(-)}(0^-)S_{\mH}\nonumber\\ -& \frac{1}{\pi}\mathcal{S}_{\mH}^{(c)}\frac{\partial\av{p}}{\partial x_1}+\frac{1}{\pi}\av{p}(0^-)S_{\mH}+\av{p} - \frac{\delta_3}{2\pi}\mathcal{S}_{\mH}^{(c)}\frac{\partial\jump{p}}{\partial x_1} + \frac{\delta_3}{2\pi}\jump{p}(0^-)S_{\mH}, \quad x_1 > 0.
\end{align}
Each of the four formulations have advantages for numerical computations depending on 
the mechanical parameters of the problem and which quantities are known or unknown. The merits of alternative formulations for the analogous isotropic case have been discussed in detail in \citet{AAG4} and we refer the reader to that paper for further discussion.

\subsection{Numerical results}
\subsubsection{Results from singular integral equations}
In this section, the integral identities found previously will be used to calculate the jump in displacement over the crack and imperfect interface between two orthotropic materials. Results for finite element simulations using COMSOL will also be presented and compared to the results using the integral identity approach derived in the previous subsection. 
%

\textcolor{black}{We will present results for the displacement jump $\jump{u}$. Note for the Mode III case that for $x_1>0$, the interfacial tractions and displacement jump $\jump{u}$ are straightforwardly related via the imperfect interface transmission conditions (\ref{jumptrans}). In particular for the Mode III displacement jump, the relationship is as follows:
\begin{equation}
 \jump{u}(x_1)=\kappa\av{t}(x_1),\qquad x_1>0.
\end{equation}
Here, we only consider tractions along the crack/interface line; discussions of full radial asymptotics (for stress and displacement) and their relationship to the displacement jump can be found in \citet{Lenci,Mishuris2001,Antipov2001,Vellender3}, among others.}

For orthotropic materials, the material parameters $H_{33}$ and $\delta_3$ are given in terms of the components of the material compliance tensor, $\mathbf{S}$, in the appendix. It is possible to express $S_{44}$ and $S_{55}$ in terms of the shear moduli, $\mu_{ij}$ of the material:
\begin{equation}
S_{44}=\frac{1}{\mu_{23}},\qquad S_{55}=\frac{1}{\mu_{13}}.
\end{equation}

In our computations, the same orthotropic material will be used as material I and II. However, the axes corresponding to each axis of symmetry of the material in the lower half-plane is altered. The parameters used for the computations presented are shown in Table \ref{mutable}. The values of $\mu_{12}$ are given in Table \ref{mutable} to illustrate that the materials considered are the same but differently oriented. Henceforth, the material above the crack (I) will be material A from Table \ref{mutable}.
\begin{table}[ht]
\begin{center}
\begin{tabular}{| l | c | c | c |}
\hline
Orientation & $\mu_{23}$ & $\mu_{13}$ & $\mu_{12}$ \\ \hline
A & 1 & 2/3 & 1/2 \\ \hline
B & 1 & 1/2 & 2/3 \\ \hline
C & 1/2 & 2/3 & 1 \\ \hline
\end{tabular}
\caption{\footnotesize Material properties}
\label{mutable}
\end{center}
\end{table}

\begin{figure}[ht]
\begin{center}
\begin{minipage}{0.9\linewidth}
	\hspace{10mm}
	\includegraphics[width=0.85\linewidth]{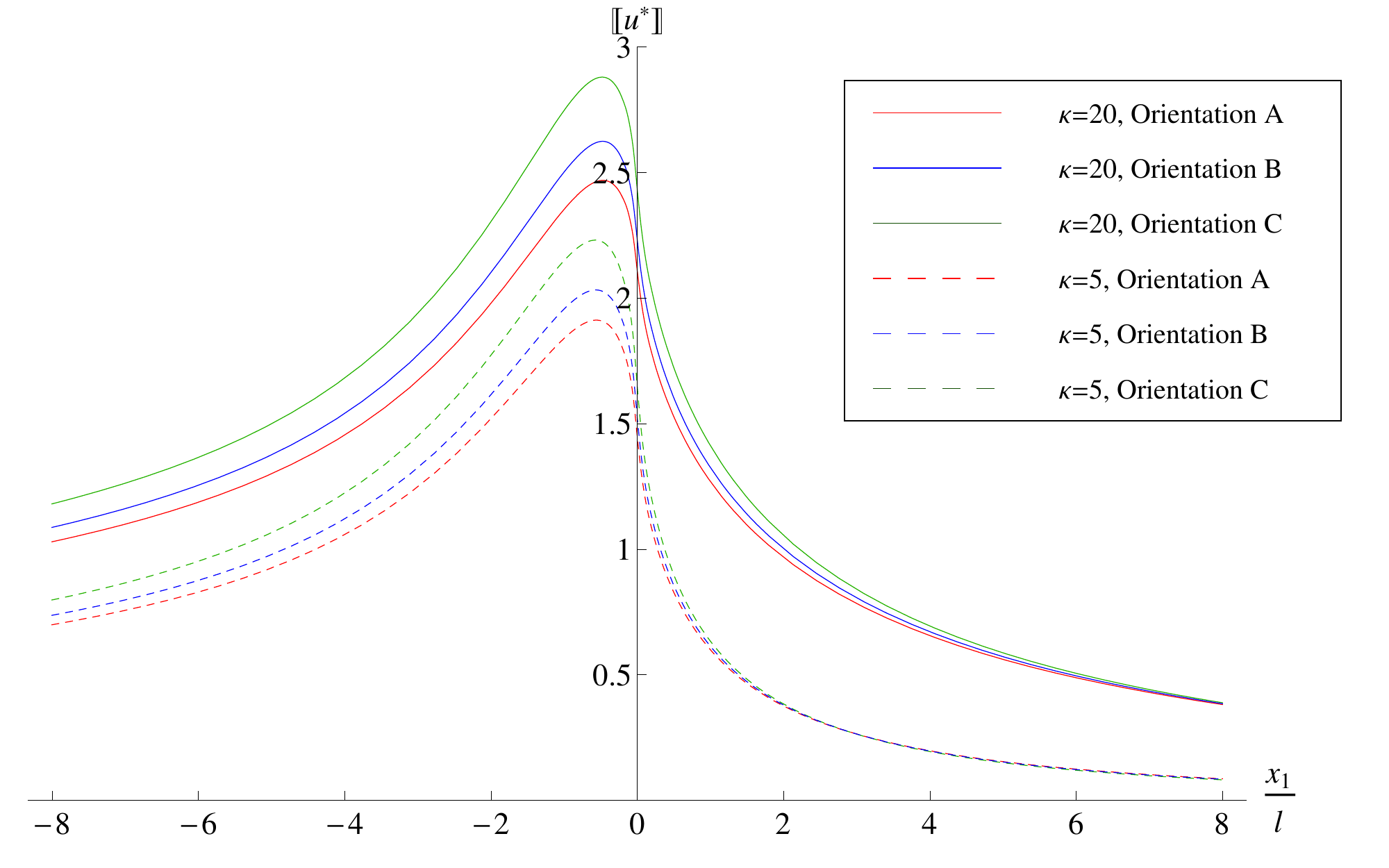}
\end{minipage}
\caption{\footnotesize Graph of normalised displacement jump over the crack and interface induced by loading (\ref{loading})\label{mathematicajump}.}
\end{center}
\end{figure}

We first consider a symmetric distribution of loadings given by
\begin{equation}\label{loading}
\jump{p}(x_1)=0,\qquad \av{p}(x_1)=-\frac{F}{l}e^{\frac{x_1}{l}}.
\end{equation}
Figure \ref{mathematicajump} plots the normalised displacement jump along the $x_1$-axis induced by the above loading for the three possible orientations for material II for two different degrees of interface imperfection  which have been computed by numerically solving the integral equations (\ref{res3b}) and (\ref{res4b}) using an iterative scheme in Mathematica. The normalised displacement jump is denoted $\jump{u^\ast}$ and defined by
\begin{equation}
\jump{u^*}=\frac{1}{F\left[ \sqrt{S_{44}S_{55}}\right]_I }\jump{u}.
\end{equation}
A normalised traction, $t^*$, is also used in the calculations and is related to the normalised displacement jump by the relationship $\jump{u^*}=\kappa^* t^*$, where
\begin{equation}
t^*=\frac{l}{F}t,\qquad \kappa^*=\frac{1}{l\left[ \sqrt{S_{44}S_{55}}\right]_I}\kappa.
\end{equation}

Figure \ref{mathematicajump} shows that a higher value of $\kappa$ gives a higher jump in displacement across the crack and interface for all orientations of the material II; this result is expected as a larger $\kappa$ refers to a less stiff interface. It is also seen that for the same value of $\kappa$, the orientation of the anisotropy has a diminishing effect along the interface ($x_1>0$) as the distance from the crack tip is increased.


The difference in orientation of material II has a clear effect on the jumps in displacement shown in Figure \ref{mathematicajump}, with the same behaviour observed for both values of $\kappa$ studied here. The highest jump in both cases is seen for orientation C in the lower half-plane. This is due to the lower shear moduli contributing to the mode-III fields in this case. 
Orientation A leads to the smallest displacement jump; this is due to the higher shear moduli in the out-of-plane direction.

\begin{figure}[ht]
\begin{center}
\begin{minipage}{0.9\linewidth}
	\hspace{10mm}
	\includegraphics[width=0.85\linewidth]{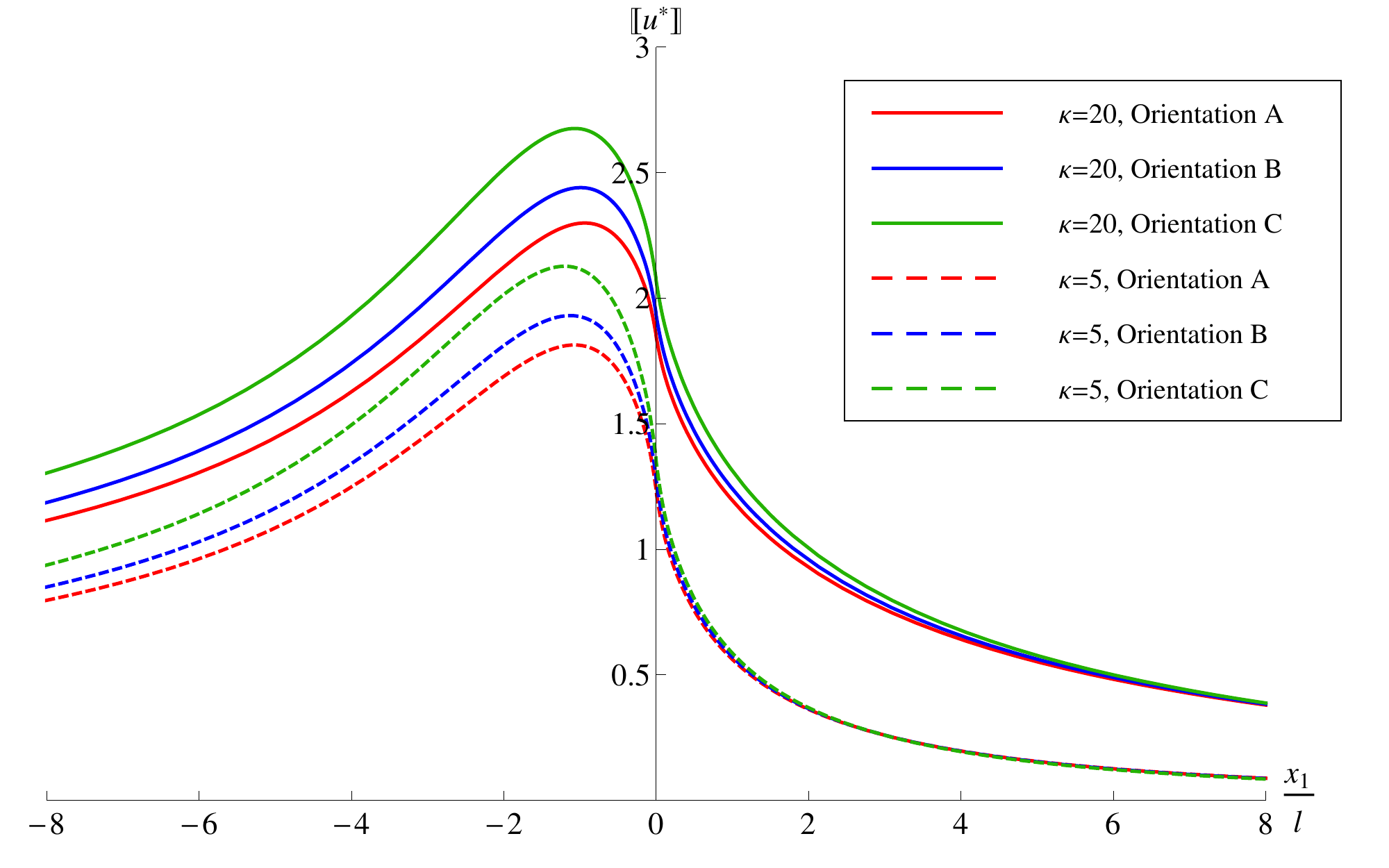}
\end{minipage}
\caption{\footnotesize Displacement jump for asymmetric loading.}\label{asymdispjump}
\end{center}
\end{figure}

In order to demonstrate that the method is applicable for asymmetric as well as symmetric loadings, we present in Figure \ref{asymdispjump} a similar plot, but instead using asymmetric loadings of the form
\begin{equation}
p^+(x_1)=-\frac{F}{l}e^{x_1/l},\qquad p^-(x_1)=\frac{F}{l^2}x_1 e^{x_1/l}.
\end{equation}

\subsubsection{Finite element results}
We now compare results from finite element simulations performed in COMSOL for a crack along an imperfect interface with computations from the integral equations. When using COMSOL it is not possible to implement the transmission conditions (\ref{tractrans}) and (\ref{jumptrans}) across the interface. 
Instead, a very thin layer of a softer material is used for the interface and the properties of that material are varied to obtain the desired value for $\kappa$ (see for instance \cite{Antipov2001}). Also, it is not possible to realise an infinite geometry in COMSOL and therefore a very large, finite geometry is used as an approximation. \textcolor{black}{These issues with the finite element model demonstrate the advantage of the boundary integral formulation, since the issues of the very fine meshing required in the interface layer and the large geometries of the main material bodies are respectively replaced by imperfect interface transmission conditions and the lower dimensional nature of the boundary problem. We present results comparing the two approaches in a case where the soft interface layer is not {\em too} thin in order to demonstrate the comparability of the two approaches.}

\begin{figure}[ht]
\begin{center}
\begin{minipage}{0.9\linewidth}
	\hspace{10mm}
	\includegraphics[width=0.85\linewidth]{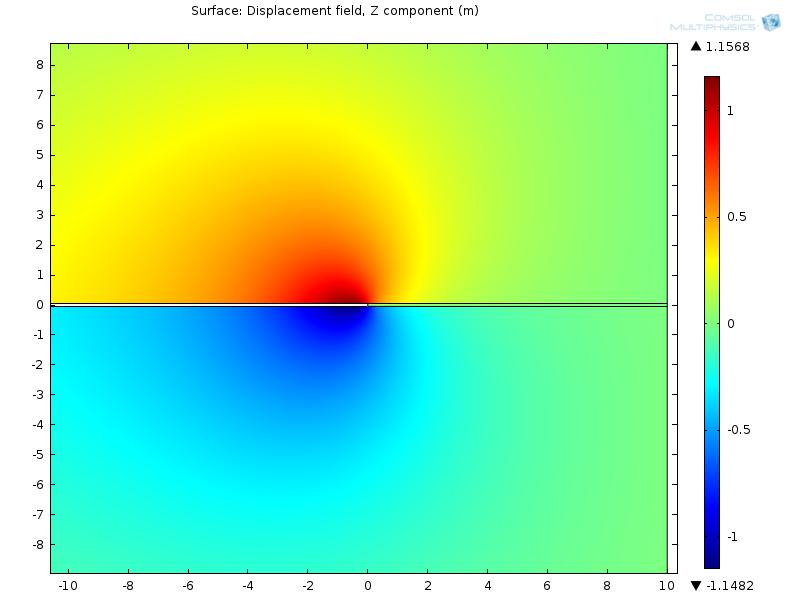}
\end{minipage}
\caption{\footnotesize Finite element computations of displacement jump, using a thin densely-meshed soft layer in place of the imperfect interface.}\label{COMSOLkappa20AA}
\end{center}
\end{figure}

An example colour map of the Mode-III displacement from COMSOL is shown in Figure \ref{COMSOLkappa20AA}, using material orientation A for both main material bodies and an interface layer corresponding to $\kappa=20$.

\begin{figure}[ht]
\begin{center}
\begin{minipage}{0.9\linewidth}
	\hspace{10mm}
	\includegraphics[width=0.85\linewidth]{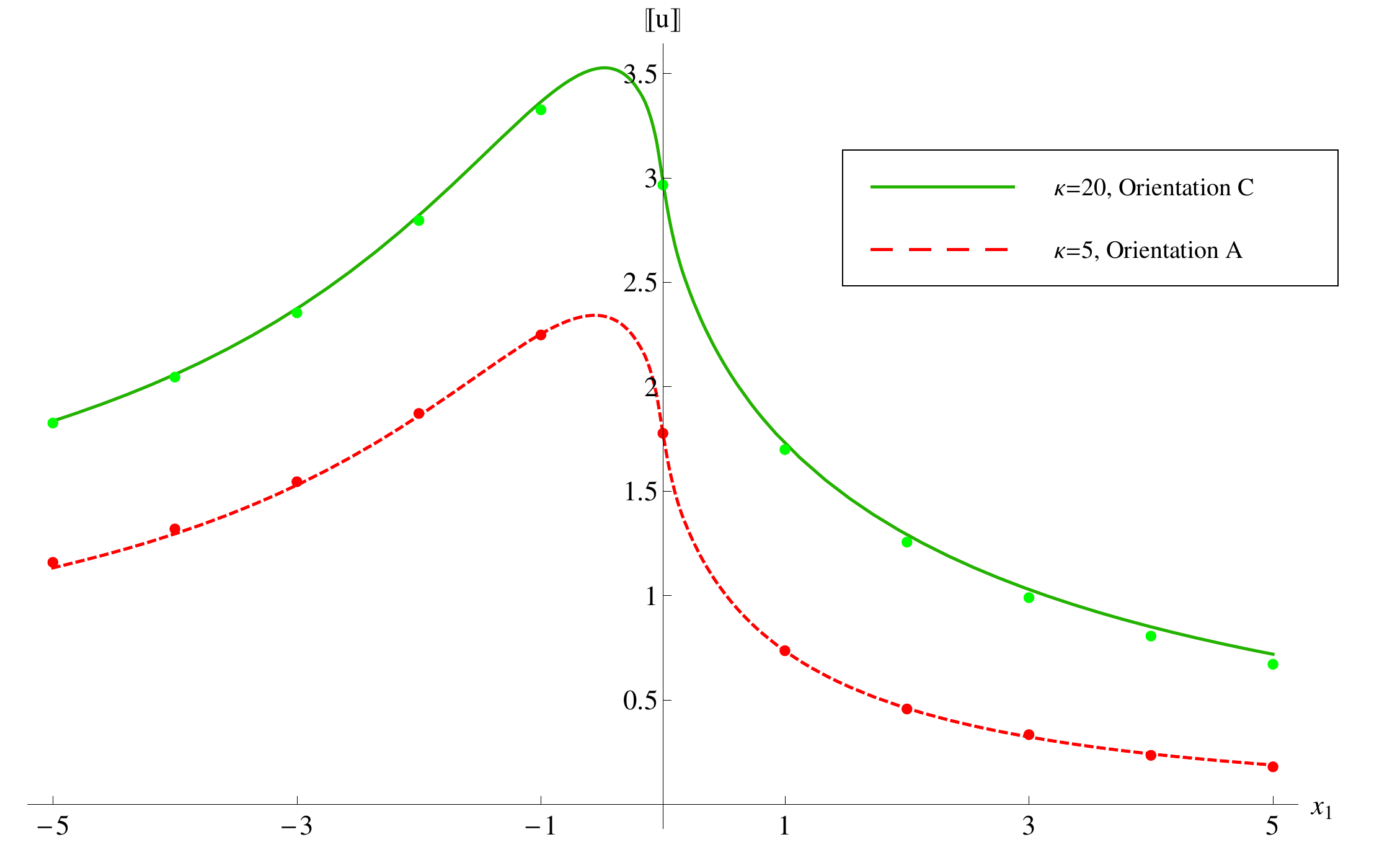}
\end{minipage}
\caption{\footnotesize Graph of the comparison between displacement jumps from Mathematica and COMSOL. The lines show the results of computations from the integral equations while finite element computations are represented by dots.}\label{compare}
\end{center}
\end{figure}

Using COMSOL, values for the displacement jump over the crack and interface have been extracted for a number of points near the crack tip for two of the examples shown in Figure \ref{mathematicajump}. 
The results of these comparisons are shown in Figure \ref{compare} and Table \ref{comparetable}.
\begin{table}[ht]
\begin{center}
\begin{tabular}{| l | c | c | c |c | c | c |c | c | c |c | c |}
\hline
Material & -5 & -4 & -3 & -2 & -1 & 0 & 1 & 2 & 3 & 4 & 5 \\ \hline
A, $\kappa=5$ & 2.30 & 1.81 & 1.07 & 0.61 & 0.20 & 0.06 & 0.18 & 0.70 & 3.41 & 2.77 & 4.66 \\ \hline
C, $\kappa=20$ & 0.53 & 0.62 & 0.84 & 0.87 & 1.13 & 0.55 & 1.80 & 2.75 & 3.81 & 5.19 & 6.70 \\ \hline
\end{tabular}
\caption{\footnotesize Percentage difference between Mathematica and COMSOL.}
\label{comparetable}
\end{center}
\end{table}

Figure \ref{compare} shows good agreement between the results from the singular integral equations and those obtained from finite element methods. The difference in results is smallest at the crack tip but more error can be seen at a further distance along both the crack and interface, which is emphasised by the larger percentage errors shown in Table \ref{comparetable}. This is likely caused by the finite geometry that was used in COMSOL which leads to an influence caused by the outer boundaries. 

%

\section{Integral identities for mode I and II}
\subsection{\textcolor{black}{Derivation of integral identities}}
Heretofore, we have derived integral identities for the mode III regime only. This section seeks to find boundary integral equations relating the mode I and II interfacial traction and displacement jump over the crack in an imperfectly bound anisotropic bimaterial. 
For the mode I and II components the following equation holds
\begin{equation}
\bar{\jump{\mathbf{U}}}^T \boldsymbol{\mathcal{R}}\overline{\av{\mathbf{t}}^{(+)}} - \overline{\av{\mathbf{\Sigma}}^{(-)}}^T\boldsymbol{\mathcal{R}}\bar{\jump{\mathbf{u}}}=-\bar{\jump{\mathbf{U}}}^T\boldsymbol{\mathcal{R}}\bar{\av{\mathbf{p}}} - \bar{\av{\mathbf{U}}}^T\boldsymbol{\mathcal{R}}\bar{\jump{\mathbf{p}}}.
\end{equation}
The matrices and vectors shown here contain only the mode I and II components from \eqref{finalbetti}. The $2\times 2$ matrices $\bar{\mathbf{U}}$ and $\bar{\mathbf{\Sigma}}$ consist of two linearly independent weight functions \citep{Piccolroaz09}.

Splitting $\bar{\jump{\mathbf{U}}}$ into the sum of $\overline{\jump{\mathbf{U}}^{(\pm)}}$ and $\bar{\jump{\mathbf{u}}}$ into $\overline{\jump{\mathbf{u}}^{(\pm)}}$, where (as previously) superscripts $^{(\pm)}$ denote the restriction of the preceding function to the respective semi-$x_1$-axis, gives
\begin{align}
\overline{\jump{\mathbf{U}}^{(+)}}^T \boldsymbol{\mathcal{R}}\overline{\av{\mathbf{t}}^{(+)}} + \overline{\jump{\mathbf{U}}^{(-)}}^T \boldsymbol{\mathcal{R}}\overline{\av{\mathbf{t}}^{(+)}} - \overline{\av{\mathbf{\Sigma}}^{(-)}}^T\boldsymbol{\mathcal{R}}\overline{\jump{\mathbf{u}}^{(+)}} &- \overline{\av{\mathbf{\Sigma}}^{(-)}}^T\boldsymbol{\mathcal{R}}\overline{\jump{\mathbf{u}}^{(-)}} \nonumber\\&= -\bar{\jump{\mathbf{U}}}^T \boldsymbol{\mathcal{R}}\bar{\av{\mathbf{p}}} - \bar{\av{\mathbf{U}}}^T\boldsymbol{\mathcal{R}}\bar{\jump{\mathbf{p}}}.
\end{align}
Applying the boundary conditions, $\overline{\jump{\mathbf{U}}^{(-)}}=0$ and $\overline{\jump{\mathbf{u}}^{(+)}}=\mathbf{K}\overline{\av{\mathbf{t}}^{(+)}}$, along with equations \eqref{weightsymm} and \eqref{weightantisymm} gives the following expression:
\begin{equation}\label{ftsol12}
\overline{\av{\mathbf{t}}^{(+)}} - \mathbf{B}(\xi)\frac{\xi}{i}\overline{\jump{\mathbf{u}}^{(-)}} = -\mathbf{C}(\xi)\bar{\av{\mathbf{p}}} - \mathbf{A}(\xi)\bar{\jump{\mathbf{p}}},
\end{equation}
where
\[
\mathbf{A}(\xi)=\frac{1}{2}\boldsymbol{\mathcal{R}}^{-1}(|\xi|\mathbf{K}^*+\mathbf{R_H}-i\mathrm{sign}(\xi)\mathbf{I_H})^{-T} (\mathbf{R_W}-i\mathrm{sign}(\xi)\mathbf{I_W})^T\boldsymbol{\mathcal{R}},
\]
\[
\mathbf{B}(\xi) = -i\boldsymbol{\mathcal{R}}^{-1}(\xi\mathbf{K}^*+\mathrm{sign}(\xi)\mathbf{R_H}-i\mathbf{I_H})^{-T}\boldsymbol{\mathcal{R}},
\]
\[
\mathbf{C}(\xi)=\boldsymbol{\mathcal{R}}^{-1}(|\xi|\mathbf{K}^*+\mathbf{R_H}-i\mathrm{sign}(\xi)\mathbf{I_H})^{-T} (\mathbf{R_H}-i\mathrm{sign}(\xi)\mathbf{I_H})^T\boldsymbol{\mathcal{R}}.
\]
Here, $\mathbf{R_{H}}=\mathrm{Re}(\mathbf{H})$, $\mathbf{R_{W}}=\mathrm{Re}(\mathbf{W})$, $\mathbf{I_{H}}=\mathrm{Im}(\mathbf{H})$, $\mathbf{I_{W}}=\mathrm{Im}(\mathbf{W})$ and $\mathbf{K}^*=\boldsymbol{\mathcal{R}}\mathbf{K}\boldsymbol{\mathcal{R}}$. Full expressions for matrices $\mathbf{A}(\xi), \mathbf{B}(\xi)$ and $\mathbf{C}(\xi)$ can be found in the appendix.

Applying the inverse Fourier transform to equation \eqref{ftsol12} for the two cases, $x_1<0$ and $x_1>0$, the following relationships are obtained:
\begin{equation}
\mathcal{F}^{-1}_{x_1<0}\left[\mathbf{B}(\xi)\frac{\xi}{i}\overline{\jump{\mathbf{u}}^{(-)}}\right] = \mathcal{F}^{-1}_{x_1<0}\left[\mathbf{C}(\xi)\bar{\av{\mathbf{p}}}\right] + \mathcal{F}^{-1}_{x_1<0}\left[\mathbf{A}(\xi)\bar{\jump{\mathbf{p}}}\right];
\end{equation}
\begin{equation}
\av{\mathbf{t}}(x_1)= \mathcal{F}^{-1}_{x_1>0}\left[\mathbf{B}(\xi)\frac{\xi}{i}\overline{\jump{\mathbf{u}}^{(-)}}\right] - \mathcal{F}^{-1}_{x_1>0}\left[\mathbf{C}(\xi)\bar{\av{\mathbf{p}}}\right] - \mathcal{F}^{-1}_{x_1>0}\left[\mathbf{A}(\xi)\bar{\jump{\mathbf{p}}}\right].
\end{equation}
The inverse Fourier transforms of the matrices $\mathbf{A}(\xi)$, $\mathbf{B}(\xi)$ and $\mathbf{C}(\xi)$ are derived in the appendix of this paper.
The singular integral equations obtained for the in-plane fields are thus
\begin{align}
\boldsymbol{\mathcal{B}}^{(s)}\frac{\partial\jump{\mathbf{u}}^{(-)}}{\partial x_1} + \frac{1}{\pi d_2(\xi_2-\xi_1)} \sum_{j = 1}^{2} \mathbf{B}_R^{(j)} T_{\xi_j}(x_1)&\jump{\mathbf{u}}^{(-)}(0^-) + \frac{1}{\pi d_2(\xi_2-\xi_1)}\sum_{j = 1}^{2} \mathbf{B}_I^{(j)} S_{\xi_j}(x_1)\jump{\mathbf{u}}^{(-)}(0^-)\nonumber\\
&= \boldsymbol{\mathcal{C}}^{(s)}\av{\mathbf{p}}(x_1) + \boldsymbol{\mathcal{A}}^{(s)}\jump{\mathbf{p}}(x_1),\quad\text{for }x_1<0,
\label{m12leftfin}\end{align}
\begin{align}
\av{\mathbf{t}}(x_1)= &\boldsymbol{\mathcal{B}}^{(c)}\frac{\partial\jump{\mathbf{u}}^{(-)}}{\partial x_1} + \frac{1}{\pi d_2(\xi_2-\xi_1)} \sum_{j = 1}^{2} \mathbf{B}_R^{(j)} T_{\xi_j}(x_1)\jump{\mathbf{u}}^{(-)}(0^-) \nonumber\\ &+\frac{1}{\pi d_2(\xi_2-\xi_1)}\sum_{j = 1}^{2} \mathbf{B}_I^{(j)} S_{\xi_j}(x_1)\jump{\mathbf{u}}^{(-)}(0^-)
-\boldsymbol{\mathcal{C}}^{(c)}\av{\mathbf{p}}(x_1) - \boldsymbol{\mathcal{A}}^{(c)}\jump{\mathbf{p}}(x_1),\quad\text{for }x_1>0.
\label{m12rightfin}\end{align}
The operators used in equations \eqref{m12leftfin} and \eqref{m12rightfin} are given by
\begin{equation}
\boldsymbol{\mathcal{A}}^{(s,c)}=-\frac{1}{2\pi d_2(\xi_2 - \xi_1)} \left\{ \sum_{j = 1}^{2} \mathbf{A}_R^{(j)} \mathcal{T}^{(s,c)}_{\xi_j}(x_1) + \sum_{j = 1}^{2} \mathbf{A}_I^{(j)} \mathcal{S}^{(s,c)}_{\xi_j}(x_1) \right\},
\end{equation}
\begin{equation}
\boldsymbol{\mathcal{B}}^{(s,c)}=-\frac{1}{\pi d_2(\xi_2 - \xi_1)} \left\{ \sum_{j = 1}^{2} \mathbf{B}_R^{(j)} \mathcal{T}^{(s,c)}_{\xi_j}(x_1) + \sum_{j = 1}^{2} \mathbf{B}_I^{(j)} \mathcal{S}^{(s,c)}_{\xi_j}(x_1) \right\},
\end{equation}
\begin{equation}
\boldsymbol{\mathcal{C}}^{(s,c)}=-\frac{1}{\pi d_2(\xi_2 - \xi_1)} \left\{ \sum_{j = 1}^{2} \mathbf{C}_R^{(j)} \mathcal{T}^{(s,c)}_{\xi_j}(x_1) + \sum_{j = 1}^{2} \mathbf{C}_I^{(j)} \mathcal{S}^{(s,c)}_{\xi_j}(x_1) \right\}.
\end{equation}

\textcolor{black}{
\subsection{Numerical examples}
In this section we present an illustrative example of applying the derived integral equations \eqref{m12leftfin} and \eqref{m12rightfin} to find the in-plane tractions and displacement jump when an asymmetrical, mode I loading is applied to the crack faces. For the purpose of these calculations, incompressible orthotropic materials will be used. It was shown by \citet{Incompressible} that for such materials only four parameters are required to express the components of $\mathbf{S}$, which are related to the matrices $\mathbf{H}$ and $\mathbf{W}$ (as seen in Appendix A). The components are 
\begin{equation}
S_{11}=\frac{1}{E_1},\quad S_{22}=\frac{1}{E_2},\quad S_{66}=\frac{1}{\mu_{12}},\nonumber
\end{equation}
\begin{equation}
S_{12}=\frac{1}{2}\left(\frac{1}{E_3}-\frac{1}{E_1}-\frac{1}{E_2}\right),
\end{equation}
where $E_i$ are the Young's moduli of the material in question. The materials considered here will have the properties shown in Table \ref{m12parameters}.}
\begin{table}[ht]
\begin{center}
\begin{tabular}{| l | c | c | c |c | }
\hline
Material & $E_1$ & $E_2$ & $E_3$ & $\mu_{12}$  \\ \hline
I & 20 & 10 & 10 & 5  \\ \hline
II& 20 & 10 & 15 & 5  \\ \hline
\end{tabular}
\caption{\footnotesize Material parameters.}
\label{m12parameters}
\end{center}
\end{table}

\begin{figure}[ht!]
\begin{center}
\begin{minipage}{\linewidth}
	\includegraphics[width=0.58\linewidth]{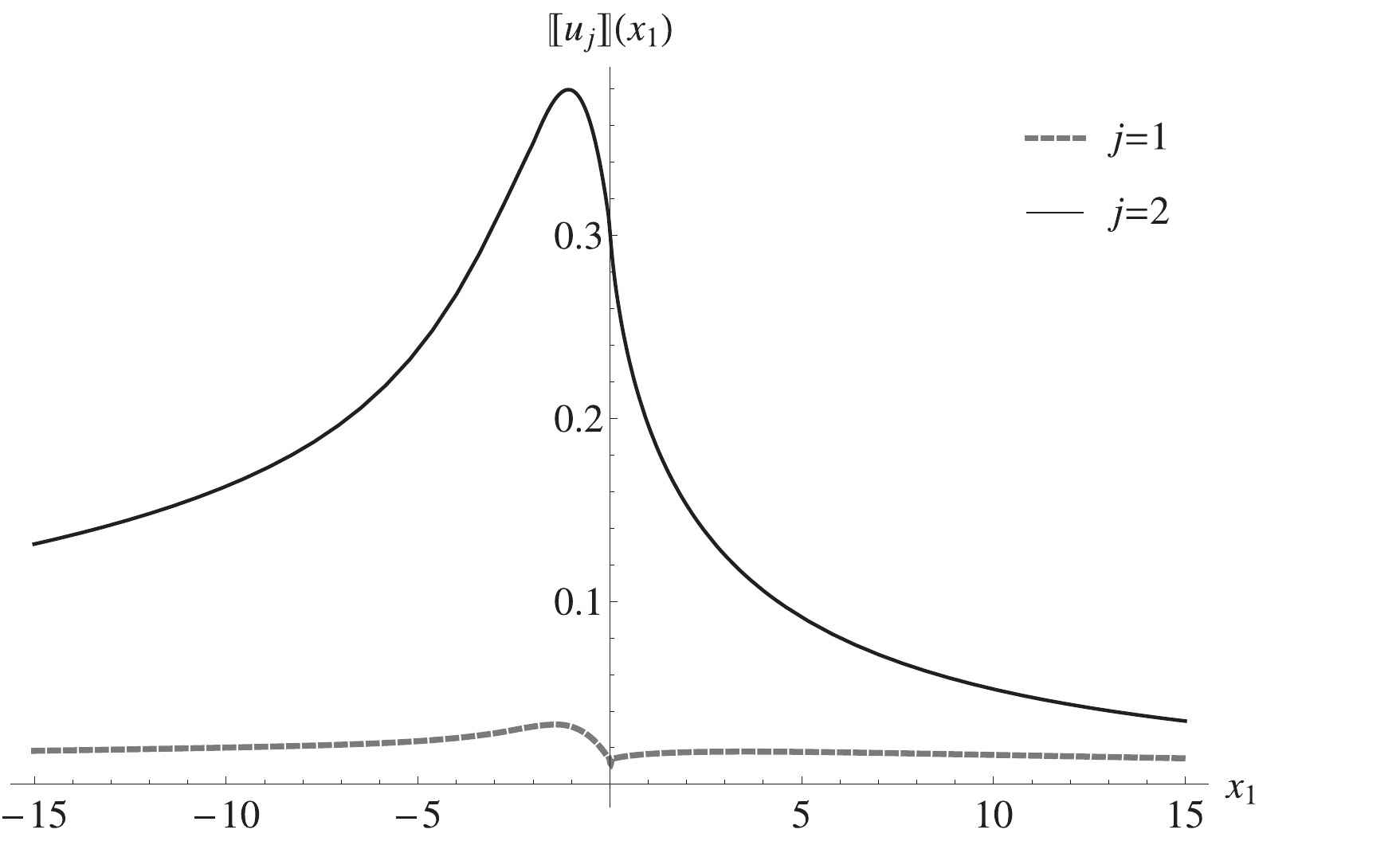}\includegraphics[width=0.42\linewidth]{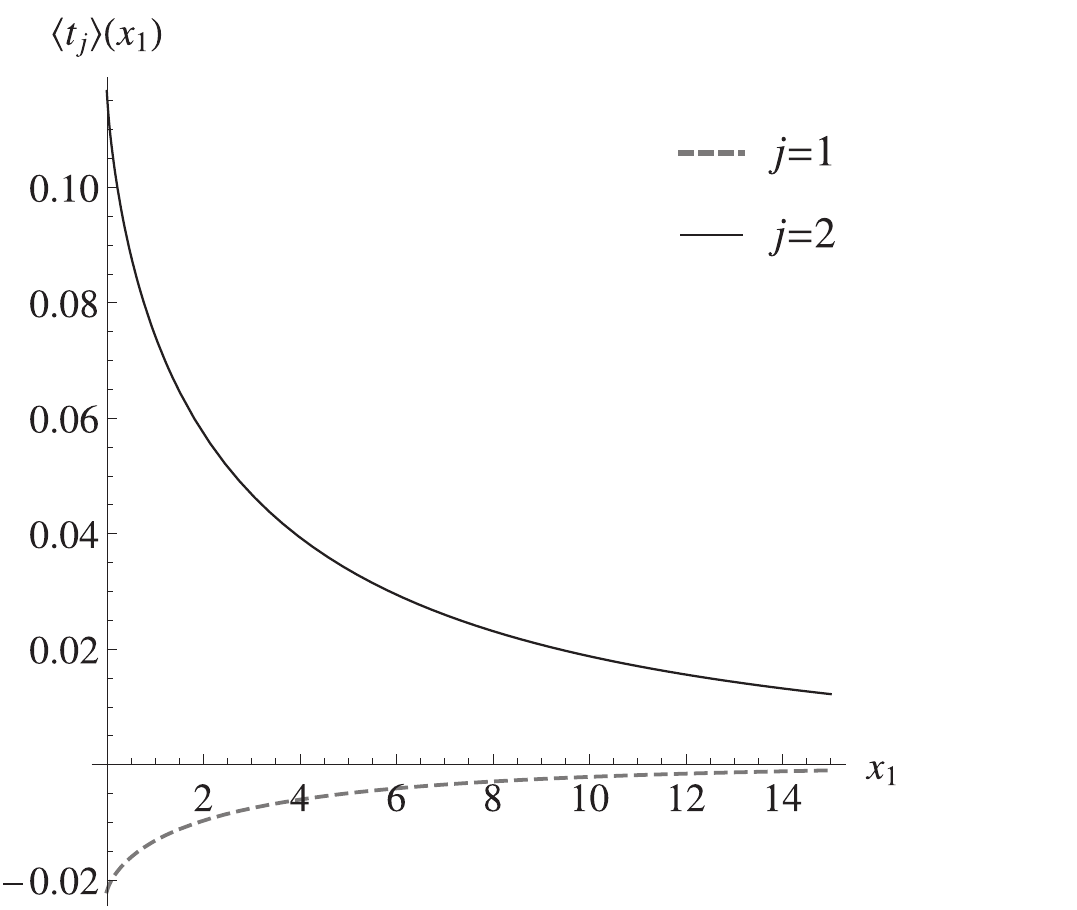}
\end{minipage}
\caption{\footnotesize \textcolor{black}{In-plane displacement jump across the crack and interface line (left), and interfacial stresses for $x_1>0$ (right).}\label{mode12}}
\end{center}
\end{figure}

\noindent\textcolor{black}{We present computations resulting from an applied asymmetric crack face loading of the form
\begin{equation}
\mathbf{p}^+(x_1)=\begin{pmatrix}0\\-\frac{F}{l}e^{x_1/l}\end{pmatrix},\quad
\mathbf{p}^-(x_1)=\begin{pmatrix}0\\\frac{F}{l^2}x_1e^{x_1/l}\end{pmatrix},
\end{equation}
with $F=1$ and $l=1$; the interfacial imperfection parameters are $K_{11}=10$, $K_{12}=2$, $K_{22}=3$. The interfacial tractions are shown in Figure \ref{mode12}, along with the displacement jump in the $x_1$ and $x_2$ directions. Note that since the crack face loadings were applied in the $x_2$-direction, the displacement jump across the crack and interface, as well as the interfacial traction, is dominant in that direction. 
Note in particular that the presence of the imperfect interface causes components of stress to remain bounded at the crack tip along the interface/crack line, in contrast to the analogous perfect interface problem.
}



\section{Conclusions}
Singular integral equations have been derived which relate the loading on crack faces to the consequent crack opening displacement and interfacial tractions for a semi-infinite crack situated along a soft anisotropic imperfect interface for an anisotropic bimaterial. The derivation made efficient use of perfect interface weight functions applied to an imperfect interface physical problem; this did not require derivation of new weight functions. As in the previously studied analogous isotropic problem, the imperfect interface's presence causes a logarithmic singularity in the kernel of the integral operator. Alternative  formulations have been presented for the mode III case and used to perform computations for orthotropic materials, 
which display a good degree of accuracy when compared against finite element simulations.

\section*{Acknowledgments}
The authors would like to thank Prof. Gennady Mishuris for fruitful discussions. LP and AZ acknowledge support from the FP7 IAPP project `INTERCER2', project reference PIAP-GA-2011-286110-INTERCER2.  AV acknowledges support from the FP7 IAPP project `HYDROFRAC', project reference PIAP-GA-2009-251475-HYDROFRAC.


\section*{References}
\begin{thebibliography}{36}
\expandafter\ifx\csname natexlab\endcsname\relax\def\natexlab#1{#1}\fi
\expandafter\ifx\csname url\endcsname\relax
  \def\url#1{\texttt{#1}}\fi
\expandafter\ifx\csname urlprefix\endcsname\relax\def\urlprefix{URL }\fi

\bibitem[{Antipov et~al.(2001)Antipov, Avila-Pozos, Kolaczkowski, and
  Movchan}]{Antipov2001}
Antipov, Y.~A., Avila-Pozos, O., Kolaczkowski, S.~T., Movchan, A.~B., 2001.
  Mathematical model of delamination cracks on imperfect interfaces. Int. J.
  Solids Struct. 38(36-37), 6665--6697.

\bibitem[{Atkinson(1977)}]{Atkinson}
Atkinson, C., 1977. On stress singularities and interfaces in linear elastic
  fracture mechanics. Int. J. Fracture 13, 807--820.

\bibitem[{Benveniste(2006)}]{Benveniste}
Benveniste, Y., 2006. A general interface model for a three-dimensional curved
  thin anisotropic interphase between two anisotropic media. J. Mech. Phys.
  Solids 54(4), 708--734.

\bibitem[{Benveniste and Miloh(2001)}]{BenvenisteMiloh}
Benveniste, Y., Miloh, T., 2001. Imperfect soft and stiff interfaces in
  two-dimensional elasticity. Mech. Materials 33, 309--323.

\bibitem[{Bueckner(1985)}]{Bueckner1}
Bueckner, H.~F., 1985. Weight functions and fundamental fields for the
  penny-shaped and the half plane crack in three-space. Int. J. Solids Struct.
  23, 57--93.

\bibitem[{Cottrell(1962)}]{Cottrell}
Cottrell, A.~H., 1962. Theoretical aspects of radiation damage and brittle
  fracture in steel pressure vessels. Iron Steel Institute Special Report 69,
  281--296.

\bibitem[{Duduchava(1979)}]{Duduchava}
Duduchava, R., 1979. Integral equations with fixed singularities. Teubner,
  Leipzig.

\bibitem[{Fett et~al.({1996})Fett, Diegele, Munz, and Rizzi}]{Fett}
Fett, T., Diegele, E., Munz, D., Rizzi, G., {1996}. {Weight functions for edge
  cracks in thin surface layers}. {Int. J. Fract.} {81}~({3}), {205--215}.

\bibitem[{Gohberg and Krein(1960)}]{GohbergKrein}
Gohberg, I.~C., Krein, M.~G., 1960. Systems of integral equations on a half
  line with kernels depending on the difference of arguments (english
  translation). Amer. Math. Soc. Transl. 14, 217--287.
  
\bibitem[{Itskov and Aksel (2002)}]{Incompressible}
Itskov, M., Aksel, N., 2002. Elastic constants and their admissible values
for incompressible and slightly compressible
anisotropic materials. Acta Mechanica. 157, 81--96.

\bibitem[{Kanninen et~al.(1979)Kanninen, Rybicki, Stonesifer, Broek,
  Rosenfiels, Marschall, and Hahn}]{Kanninen}
Kanninen, M.~F., Rybicki, E.~F., Stonesifer, R.~B., Broek, D., Rosenfiels,
  A.~R., Marschall, C.~W., Hahn, G.~T., 1979. Elastic-plastic fracture
  mechanics for two dimensional stable crack growth and instability problems.
  Elastic-Plastic Fracture ASTM STP 668, 121--150.

\bibitem[{Kassir and Sih(1973)}]{KassirSih}
Kassir, M.~K., Sih, G.~C., 1973. Application of papkovich-neuber potentials to
  a crack problem. Int. J. Solids Struct. 9, 643--654.

\bibitem[{Lekhnitskii(1963)}]{Lek}
Lekhnitskii, S.~G., 1963. Theory of Elasticity of an Anisotropic Body. MIR,
  Moscow.

\bibitem[{Lenci(2001)}]{Lenci}
Lenci, S., 2001. Analysis of a crack at a weak interface. Int. J. Fract. 108,
  275--290.

\bibitem[{Mishuris(2001)}]{Mishuris2001}
Mishuris, G., 2001. Interface crack and nonideal interface concept (mode iii).
  Int. J. Fract. 107(3), 279--296.

\bibitem[{Mishuris and Kuhn(2001)}]{MisKuhn}
Mishuris, G., Kuhn, G., 2001. Asymptotic behaviour of the elastic solution near
  the tip of a crack situated at a nonideal interface. Zeitschrift fur
  Angewandte Mathematik und Mechanik 81(12), 811--826.

\bibitem[{Mishuris et~al.(2014)Mishuris, Piccolroaz, and Vellender}]{AAG4}
Mishuris, G., Piccolroaz, A., Vellender, A., 2014. Boundary integral
  formulation for cracks at imperfect interfaces. Q. J. Mech. Appl. Math.
  DOI:10.1093/qjmam/hbu010 (Available online).

\bibitem[{Mishuris(1997{\natexlab{a}})}]{Mishuris1997a}
Mishuris, G.~S., 1997{\natexlab{a}}. 2-d boundary value problems of
  thermoelasticity in a multi-wedge -- multi-layered region. part 1. sweep
  method. Arch. Mech. 49(6), 1103--1134.

\bibitem[{Mishuris(1997{\natexlab{b}})}]{Mishuris1997b}
Mishuris, G.~S., 1997{\natexlab{b}}. 2-d boundary value problems of
  thermoelasticity in a multi-wedge -- multi-layered region. part 2. systems of
  integral equations. Arch. Mech. 49(6), 1135--1165.

\bibitem[{Morini et~al.(2013{\natexlab{a}})Morini, Piccolroaz, Mishuris, and
  Radi}]{Morini2}
Morini, L., Piccolroaz, A., Mishuris, G., Radi, E., 2013{\natexlab{a}}.
  Integral identities for a semi-infinite interfacial crack in anisotropic
  elastic bimaterials. Int. J. Solids Struct. 50, 1437--1448.

\bibitem[{Morini et~al.(2013{\natexlab{b}})Morini, Radi, Movchan, and
  Movchan}]{Morini}
Morini, L., Radi, E., Movchan, A.~B., Movchan, N.~V., 2013{\natexlab{b}}. Stroh
  formalism in analysis of skew-symmetric and symmetric weight functions for
  interfacial cracks. Math. Mech. Solids 18, 135--152.

\bibitem[{Muskhelishvili(1963)}]{Mus}
Muskhelishvili, N.~I., 1963. Some Basic Problems of the Mathematical Theory of
  Elasticity. Groningen: P.Noordhoff, Netherlands.

\bibitem[{Piccolroaz and Mishuris(2013)}]{Piccolroaz13}
Piccolroaz, A., Mishuris, G., 2013. Integral identities for a semi-infinite
  interfacial crack in 2d and 3d elasticity. J. Elasticity 110, 117--140.

\bibitem[{Piccolroaz et~al.(2007)Piccolroaz, Mishuris, and
  Movchan}]{Piccolroaz07}
Piccolroaz, A., Mishuris, G., Movchan, A.~B., 2007. Evaluation of the
  lazarus-leblond constants in the asymptotic model for the interfacial wavy
  crack. J. Mech. Phys. Solids 55, 1575--1600.

\bibitem[{Piccolroaz et~al.(2009)Piccolroaz, Mishuris, and
  Movchan}]{Piccolroaz09}
Piccolroaz, A., Mishuris, G., Movchan, A.~B., 2009. Symmetric and
  skew-symmetric weight functions in 2d perturbation models for semi-infinite
  interfacial cracks. J. Mech. Phys. Solids 57, 1657--1682.

\bibitem[{Pryce et~al.(2013)Pryce, Morini, and Mishuris}]{Pryce1}
Pryce, L., Morini, L., Mishuris, G., 2013. Weight function approach to study a
  crack propagating along a bimaterial interface under arbitrary loading in
  anisotropic solids. JoMMS 8, 479--500.

\bibitem[{Rice and Sorenson(1978)}]{RiceSorenson}
Rice, J.~R., Sorenson, E.~P., 1978. Continuing crack tip deformation and
  fracture for plane strain crack growth in elastic-plastic solids. J. Mech.
  Phys. Solids 26, 163--186.

\bibitem[{Shih et~al.(1979)Shih, de~Lorenzi, and Andrews}]{Shih}
Shih, C.~F., de~Lorenzi, H.~G., Andrews, W.~R., 1979. Studies on crack
  initiation and stable crack growth. Elastic-Plastic Fracture ASTM STP 668,
  65--120.

\bibitem[{Sneddon(1972)}]{Sneddon}
Sneddon, I.~N., 1972. The use of integral transforms. McGraw-Hill, New York.

\bibitem[{Stroh(1962)}]{Stroh}
Stroh, A.~N., 1962. Steady state problems in anisotropic elasticity. Math. Phys
  41, 77--103.

\bibitem[{Suo(1990)}]{Suo}
Suo, Z., 1990. Singularities, interfaces and cracks in dissimilar anisotropic
  media. Proc. R. Soc. Lond 427, 331--358.

\bibitem[{Vellender and Mishuris(2012)}]{Vellender2}
Vellender, A., Mishuris, G.~S., 2012. Eigenfrequency correction of
  bloch-floquet waves in a thin periodic bi-material strip with cracks lying on
  perfect and imperfect interfaces. Wave Motion 49(2), 258--270.

\bibitem[{Vellender et~al.(2011)Vellender, Mishuris, and Movchan}]{Vellender1}
Vellender, A., Mishuris, G.~S., Movchan, A.~B., 2011. Weight function in a
  bimaterial strip containing an interfacial crack and an imperfect interface.
  application to a bloch-floquet analysis in a thin inhomogeneous structure
  with cracks. Multiscale Model. Simul. 9(4), 1327--1349.

\bibitem[{Vellender et~al.(2013)Vellender, Mishuris, and
  Piccolroaz}]{Vellender3}
Vellender, A., Mishuris, G.~S., Piccolroaz, A., 2013. Perturbation analysis for
  an imperfect interface crack problem using weight function techniques. Int.
  J. Solids Struct. 50(24), 4098--4107.

\bibitem[{Wells(1961)}]{Wells}
Wells, A.~A., 1961. Unstable crack propagation in metals: Cleavage and
  fracture. Proceedings of the crack propagation symposium, Cranfield,
  210--230.

\bibitem[{Willis and Movchan(1995)}]{WillisMovchan}
Willis, J.~R., Movchan, A.~B., 1995. Dynamic weight function for a moving
  crack. i. mode i loading. J. Mech. Phys. Solids, 319--341.

\bibitem[{Yu and Suo(2000)}]{YuSuo}
Yu, H.~H., Suo, Z., 2000. Intersonic crack growth on an interface. Proc. R. Soc. Lond. 456, 223-246.
  
\bibitem[{Zheng et~al.(1996)Zheng, Glinka, and Dubey}]{ZhengGlinka}
Zheng, X.~J., Glinka, G., Dubey, R.~N., 1996. Stress intensity factors and
  weight functions for a corner crack in a ﬁnite thickness plate. Eng. Frac.
  Mech. 54(1), 49--61.

\end{thebibliography}

\appendix

\section{Bimaterial matrices $\mathbf{H}$ and $\mathbf{W}$ for orthotropic bimaterials}
The matrices $\mathbf{H}$ and $\mathbf{W}$ have the form
\begin{equation}
\mathbf{H}=\begin{pmatrix} H_{11}&-i\beta\sqrt{H_{11}H_{22}}&0\\i\beta\sqrt{H_{11}H_{22}}&H_{22}&0\\0&0&H_{33}\end{pmatrix},\quad
\mathbf{W}=\begin{pmatrix} \delta_1 H_{11}&i\gamma\sqrt{H_{11}H_{22}}&0\\-i\gamma\sqrt{H_{11}H_{22}}&\delta_2 H_{22}&0\\0&0&\delta_3 H_{33}\end{pmatrix}.
\end{equation}
For orthotropic materials it is possible to obtain explicit expressions for the these matrices in terms of the components of the material compliance tensors.

The out-of-plane components are given by
\begin{equation}
H_{33}=\left[ \sqrt{S_{44}S_{55}}\right]_I + \left[ \sqrt{S_{44}S_{55}}\right]_{II},\quad 
\delta_3 = \frac{\left[ \sqrt{S_{44}S_{55}}\right]_I - \left[ \sqrt{S_{44}S_{55}}\right]_{II}}{H_{33}}.
\end{equation}

The in-plane components of $\mathbf{H}$ can be found in \citet{Morini} and are given as
\begin{equation}
H_{11}=\left[ 2n\lambda^{1/4}\sqrt{S_{11}S_{22}}\right]_I + \left[ 2n\lambda^{1/4}\sqrt{S_{11}S_{22}}\right]_{II},\end{equation}
\begin{equation}
H_{22}=\left[ 2n\lambda^{-1/4}\sqrt{S_{11}S_{22}}\right]_I + \left[ 2n\lambda^{-1/4}\sqrt{S_{11}S_{22}}\right]_{II},
\end{equation}
\begin{equation}
\beta=\frac{\left[ S_{12}+\sqrt{S_{11}S_{22}}\right]_{II} - \left[ S_{12}+\sqrt{S_{11}S_{22}}\right]_{I}}{\sqrt{H_{11}H_{22}}},
\end{equation}
where
\[
\lambda=\frac{S_{11}}{S_{22}},\quad n=\sqrt{(1+\rho)/2},\quad \rho=\frac{2S_{12}+S_{66}}{2\sqrt{S_{11}S_{22}}}.
\]

The in-plane components of $\mathbf{W}$ were also given in \citet{Morini}:
\begin{equation}
\delta_1=\frac{\left[ 2n\lambda^{1/4}\sqrt{S_{11}S_{22}}\right]_I - \left[ 2n\lambda^{1/4}\sqrt{S_{11}S_{22}}\right]_{II}}{H_{11}},\end{equation}
\begin{equation}
\delta_2=\frac{\left[ 2n\lambda^{-1/4}\sqrt{S_{11}S_{22}}\right]_I - \left[ 2n\lambda^{-1/4}\sqrt{S_{11}S_{22}}\right]_{II}}{H_{22}},
\end{equation}
\begin{equation}
\gamma= \frac{\left[S_{12}+\sqrt{S_{11}S_{22}}\right]_I + \left[S_{12}+\sqrt{S_{11}S_{22}}\right]_{II}}{\sqrt{H_{11}H_{22}}}.
\end{equation}

\section{The matrices $\mathbf{A}(\xi)$, $\mathbf{B}(\xi)$ and $\mathbf{C}(\xi)$}
Matrices $\mathbf{A}(\xi)$, $\mathbf{B}(\xi)$ and $\mathbf{C}(\xi)$ have the following form
\begin{equation}
\mathbf{A}(\xi) = \frac{1}{2D}
\begin{pmatrix}
A_{11} & A_{12} \\
A_{21} & A_{22}
\end{pmatrix}, \quad
\mathbf{B}(\xi) = \frac{1}{D}
\begin{pmatrix}
B_{11} & B_{12} \\
B_{21} & B_{22}
\end{pmatrix}, \quad
\mathbf{C}(\xi) = \frac{1}{D}
\begin{pmatrix}
C_{11} & C_{12} \\
C_{21} & C_{22}
\end{pmatrix}
\end{equation}
where the denominator $D$ is defined as
\begin{equation}
\label{denominator}
D = d_0 + d_1 |\xi| + d_2 |\xi|^2,
\end{equation}
\[
d_0 = H_{11}H_{22}(1-\beta^2), \quad
d_1 = K_{11}H_{22}+K_{22}H_{11}, \quad
d_2 = K_{11}K_{22}-K_{12}^2,
\]
and the elements $A_{ij}$, $B_{ij}$, $C_{ij}$ are given by
\[
A_{11} =  H_{11}H_{22}(\delta_1 +\beta\gamma)+|\xi|(\delta_1H_{11}K_{22}-i\gamma K_{12}\sqrt{H_{11}H_{22}}\text{ sign}(\xi)),\]\[
A_{12} = -i\text{ sign}(\xi)H_{22}\sqrt{H_{11}H_{22}}(\gamma +\beta\delta_2)-|\xi|(i\gamma K_{22}\sqrt{H_{11}H_{22}}\text{ sign}(\xi)+\delta_2H_{22}K_{12}),\]\[
A_{21} = i\text{ sign}(\xi)H_{11}\sqrt{H_{11}H_{22}}(\delta_1\beta +\gamma)-|\xi|(\delta_1 H_{11}K_{12}-i\gamma K_{11}\sqrt{H_{11}H_{22}}\text{ sign}(\xi)),\]\[
A_{22} = H_{11}H_{22}(\beta\gamma +\delta_2)+|\xi|(\delta_2 H_{22} K_{11}+i\gamma K_{12}\sqrt{H_{11}H_{22}}\text{ sign}(\xi)),
\]
\vspace*{0.4cm}
\[
B_{11} = -i(\xi K_{22} +H_{22}\text{ sign}(\xi)),\]\[
B_{12} = i\xi K_{12}-\beta\sqrt{H_{11}H_{22}},\]\[
B_{21} = i\xi K_{12}+\beta\sqrt{H_{11}H_{22}},\]\[
B_{22} = -i(\xi K_{11} +H_{11}\text{ sign}(\xi)),
\]
\vspace*{0.4cm}
\[
C_{11} = H_{11}H_{22}(1-\beta^2)+|\xi|(H_{11}K_{22}+i\beta K_{12}\sqrt{H_{11}H_{22}}\text{ sign}(\xi)),\]\[
C_{12} = -|\xi|(H_{22}K_{12}-i\beta\text{ sign}(\xi)K_{22}\sqrt{H_{11}H_{22}}),\]\[
C_{21} = -|\xi|(H_{11}K_{12}+i\beta\text{ sign}(\xi)K_{11}\sqrt{H_{11}H_{22}}),\]\[
C_{22} = H_{11}H_{22}(1-\beta^2)+|\xi|(H_{22}K_{11}-i\beta K_{12}\sqrt{H_{11}H_{22}}\text{ sign}(\xi)).
\]

\section{Inverse Fourier transforms of matrices $\mathbf{A}(\xi)$, $\mathbf{B}(\xi)$ and $\mathbf{C}(\xi)$}
\subsection{General procedure}

The method outlined in \citet{AAG4} is used in order to perform the Fourier inversion of the matrices $\mathbf{A}(\xi)$, $\mathbf{B}(\xi)$ and $\mathbf{C}(\xi)$. The denominator $D$ defined in 
(\ref{denominator}) is factorised in the following manner
\begin{equation}
D = d_2 (|\xi| + \xi_1) (|\xi| + \xi_2),
\end{equation}
where
\begin{equation}
\label{chi12}
\xi_{1,2} = \frac{d_1 \mp \sqrt{d_1^2 - 4d_2d_0}}{2d_2} > 0,
\end{equation}
The typical term to invert is of the form
\begin{equation}
F(\xi) = \frac{F_R + F_R^\dag|\xi|}{D} + i \frac{F_I \text{ sign}(\xi) + F_I^\dag\xi}{D},
\end{equation}
The function $F$ has the following property
\begin{equation}
F(-\xi) = \overline{F(\xi)},
\end{equation}
therefore, the Fourier inversion can be obtained as
\begin{equation}
\mathcal{F}^{-1}[F(\xi)] = \frac{1}{\pi} \mathrm{Re} \int_0^\infty F(\xi) e^{-ix_1\xi} \mathrm{d}\xi = 
\frac{1}{\pi} \int_0^\infty \mathrm{Re}[F(\xi)] \cos(x_1\xi) \mathrm{d}\xi + \frac{1}{\pi} \int_0^\infty \mathrm{Im}[F(\xi)] \sin(x_1\xi) \mathrm{d}\xi,
\end{equation}
where for $\xi > 0$
\begin{equation}
\mathrm{Re}[F(\xi)] = \frac{F_{R} + F_{R}^\dag\xi}{D} = \sum_{j = 1}^2 \frac{F_{R}^{(j)}}{d_2(\xi_2 - \xi_1)(\xi + \xi_j)},
\end{equation}
\begin{equation}
\mathrm{Im}[F(\xi)] = \frac{F_{I} + F_{I}^\dag\xi}{D} = \sum_{j = 1}^2 \frac{F_{I}^{(j)}}{d_2(\xi_2 - \xi_1)(\xi + \xi_j)},
\end{equation}
and
\begin{equation}
F_{R,I}^{(1)} = F_{R,I} - F_{R,I}^\dag\xi_1, \quad F_{R,I}^{(2)} = -F_{R,I} + F_{R,I}^\dag\xi_2.
\end{equation}
The following formulae can now be used
\begin{equation}
\int_0^\infty \mathrm{Re}[F(\xi)] \cos(x_1\xi) \mathrm{d}\xi = \sum_{j = 1}^{2} \frac{F_R^{(j)}}{d_2(\xi_2 - \xi_1)} \int_0^\infty \frac{\cos(x_1\xi)}{\xi + \xi_j} \mathrm{d}\xi = 
-\frac{1}{d_2(\xi_2 - \xi_1)} \sum_{j = 1}^{2} F_R^{(j)} T_{\xi_j}(x_1),
\end{equation}
\begin{equation}
\int_0^\infty \mathrm{Im}[F(\xi)] \sin(x_1\xi) \mathrm{d}\xi = \sum_{j = 1}^{2} \frac{F_I^{(j)}}{d_2(\xi_2 - \xi_1)} \int_0^\infty \frac{\sin(x_1\xi)}{\xi + \xi_j} \mathrm{d}\xi = 
-\frac{1}{d_2(\xi_2 - \xi_1)} \sum_{j = 1}^{2} F_I^{(j)} S_{\xi_j}(x),
\end{equation}
where functions $S_{\xi_j}(x)$ and $T_{\xi_j}(x)$ are defined as in \eqref{S} and \eqref{T}, respectively.

Finally the Fourier inversion of the general term $F(\xi)$ as given as
\begin{equation}
\mathcal{F}^{-1}[F(\xi)] = -\frac{1}{\pi d_2(\xi_2 - \xi_1)} \left\{ \sum_{j = 1}^{2} F_R^{(j)} T_{\xi_j}(x_1) 
+ \sum_{j = 1}^{2} F_I^{(j)} S_{\xi_j}(x_1) \right\}.
\end{equation}

\subsection{Fourier inversion of $\mathbf{A}(\xi)$.}

For $\xi > 0$, $\mathbf{A}(\xi)$ can be written as 
\begin{equation}
\mathbf{A}(\xi) 
= \frac{1}{2D}(\mathbf{A}_R + \mathbf{A}_R^\dag \xi) + \frac{i}{2D} (\mathbf{A}_I + \mathbf{A}_I^\dag \xi) 
= \frac{1}{2d_2(\xi_2 - \xi_1)} \left\{ \sum_{j = 1}^{2} \frac{1}{\xi + \xi_j} \mathbf{A}_R^{(j)} + i \sum_{j = 1}^{2} \frac{1}{\xi + \xi_j} \mathbf{A}_I^{(j)} \right\},
\end{equation}
where
\begin{equation}
\mathbf{A}_R =H_{11}H_{22}\begin{pmatrix}\delta_1 +\beta\gamma & 0\\0 & \delta_2+\beta\gamma\end{pmatrix}, \quad
\mathbf{A}_R^\dag =
\begin{pmatrix}
\delta_1 H_{11}K_{22} & -\delta_2 H_{22}K_{12} \\
-\delta_1 H_{11}K_{12} & \delta_2 H_{22}K_{11}
\end{pmatrix},
\end{equation}
\begin{equation}
\mathbf{A}_I = \sqrt{H_{11}H_{22}}\begin{pmatrix}0 & - H_{22}(\delta_2\beta+\gamma) \\ H_{11}(\delta_1\beta +\gamma)& 0\end{pmatrix}, \quad
\mathbf{A}_I^\dag = \gamma\sqrt{H_{11}H_{22}}
\begin{pmatrix}
-K_{12} & -K_{22} \\
K_{11} & K_{12}
\end{pmatrix},
\end{equation}
\begin{equation}
\mathbf{A}_R^{(1)} = \mathbf{A}_R - \mathbf{A}_R^\dag \xi_1 = 
\begin{pmatrix}
H_{11}(H_{22}(\delta_1 +\beta\gamma)-\delta_1K_{22}\xi_1) & \delta_2H_{22}K_{12}\xi_1 \\
\delta_1H_{11}K_{12}\xi_1 & H_{22}(H_{11}(\delta_2 +\beta\gamma)-\delta_2K_{11}\xi_1)
\end{pmatrix},
\end{equation}
\begin{equation}
\mathbf{A}_R^{(2)} = -\mathbf{A}_R + \mathbf{A}_R^\dag \xi_2 =
\begin{pmatrix}
-H_{11}(H_{22}(\delta_1+\beta\gamma) - \delta_1K_{22}\xi_2) & -\delta_2H_{22}K_{12}\xi_2 \\
-\delta_1H_{11}K_{12}\xi_2 & -H_{22}(H_{11}(\delta_2+\beta\gamma) - \delta_2K_{11}\xi_2)
\end{pmatrix},
\end{equation}
\begin{equation}
\mathbf{A}_I^{(1)} = \mathbf{A}_I - \mathbf{A}_I^\dag \xi_1 = 
\sqrt{H_{11}H_{22}}\begin{pmatrix}
\gamma K_{12}\xi_1 & -H_{22}(\beta\delta_2 +\gamma )+\gamma K_{22}\xi_1 \\
H_{11}(\beta\delta_1 +\gamma)-\gamma K_{11}\xi_1 & -\gamma K_{12}\xi_1
\end{pmatrix},
\end{equation}
\begin{equation}
\mathbf{A}_I^{(2)} = -\mathbf{A}_I + \mathbf{A}_I^\dag \xi_2 =
\sqrt{H_{11}H_{22}}\begin{pmatrix}
-\gamma K_{12}\xi_2 & H_{22}(\beta\delta_2 +\gamma)-\gamma K_{22}\xi_2 \\
-H_{11}(\beta\delta_1 +\gamma)+\gamma K_{11}\xi_2 & \gamma K_{12}\xi_2
\end{pmatrix}.
\end{equation}
The Fourier inverse of the matrix $\mathbf{A}(\xi)$ is given by
\begin{equation}
\mathcal{F}^{-1}[\mathbf{A}(\xi)] = -\frac{1}{2\pi d_2(\xi_2 - \xi_1)} \left\{ \sum_{j = 1}^{2} \mathbf{A}_R^{(j)} T_{\xi_j}(x_1) + \sum_{j = 1}^{2} \mathbf{A}_I^{(j)} S_{\xi_j}(x_1) \right\}.
\end{equation}

\subsection{Fourier inversion of the matrix $\mathbf{B}(\xi)$.}

For $\xi > 0$ $\mathbf{B}(\xi)$ can be written as 
\begin{equation}
\mathbf{B}(\xi) = \frac{1}{D}(\mathbf{B}_R + \mathbf{B}_R^\dag \xi) + \frac{i}{D} (\mathbf{B}_I + \mathbf{B}_I^\dag \xi) = 
\frac{1}{d_2(\xi_2 - \xi_1)} \left\{ \sum_{j = 1}^{2} \frac{1}{\xi + \xi_j} \mathbf{B}_R^{(j)} + i \sum_{j = 1}^{2} \frac{1}{\xi + \xi_j} \mathbf{B}_I^{(j)} \right\},
\end{equation}
where
\begin{equation}
\mathbf{B}_R = \beta\sqrt{H_{11}H_{22}}\begin{pmatrix}0&-1\\ 1&0\end{pmatrix}, \quad
\mathbf{B}_R^\dag = \mathbf{0},
\end{equation}
\begin{equation}
\mathbf{B}_I = \begin{pmatrix}-H_{22}&0\\0&-H_{11}\end{pmatrix}, \quad
\mathbf{B}_I^\dag = 
\begin{pmatrix}
-K_{22} & K_{12} \\
K_{12} & -K_{11}
\end{pmatrix},
\end{equation}
\begin{equation}
\mathbf{B}_R^{(1)} = \mathbf{B}_R - \mathbf{B}_R^\dag \xi_1 = 
\beta\sqrt{H_{11}H_{22}}\begin{pmatrix}0&-1\\ 1&0\end{pmatrix},
\end{equation}
\begin{equation}
\mathbf{B}_R^{(2)} = -\mathbf{B}_R + \mathbf{B}_R^\dag \xi_2 =
\beta\sqrt{H_{11}H_{22}}\begin{pmatrix}0&1\\-1&0\end{pmatrix},
\end{equation}
\begin{equation}
\mathbf{B}_I^{(1)} = \mathbf{B}_I - \mathbf{B}_I^\dag \xi_1 = 
\begin{pmatrix}
-H_{22}+K_{22}\xi_1 & -K_{12}\xi_1 \\
-K_{12}\xi_1 & -H_{11}+K_{11}\xi_1
\end{pmatrix},
\end{equation}
\begin{equation}
\mathbf{B}_I^{(2)} = -\mathbf{B}_I + \mathbf{B}_I^\dag \xi_2 = 
\begin{pmatrix}
H_{22}-K_{22}\xi_2 & K_{12}\xi_2 \\
K_{12}\xi_2 & H_{11}-K_{11}\xi_2
\end{pmatrix}.
\end{equation}
The Fourier inverse of the matrix $\mathbf{B}(\xi)$ is then
\begin{equation}
\mathcal{F}^{-1}[\mathbf{B}(\xi)] = -\frac{1}{\pi d_2(\xi_2 - \xi_1)} \left\{ \sum_{j = 1}^{2} \mathbf{B}_R^{(j)} T_{\xi_j}(x_1) + \sum_{j = 1}^{2} \mathbf{B}_I^{(j)} S_{\xi_j}(x_1) \right\}.
\end{equation}

\subsection{Fourier inversion of the matrix $\mathbf{C}(\xi)$.}

For $\xi > 0$ $\mathbf{C}(\xi)$ can be written as
\begin{equation}
\mathbf{C}(\xi) = \frac{1}{D}(\mathbf{C}_R + \mathbf{C}_R^\dag \xi) + \frac{i}{D} (\mathbf{C}_I + \mathbf{C}_I^\dag \xi) = 
\frac{1}{d_2(\xi_2 - \xi_1)} \left\{ \sum_{j = 1}^{2} \frac{1}{\xi + \xi_j} \mathbf{C}_R^{(j)} + i \sum_{j = 1}^{2} \frac{1}{\xi + \xi_j} \mathbf{C}_I^{(j)} \right\},
\end{equation}
where
\begin{equation}
\mathbf{C}_R = \begin{pmatrix}H_{11}H_{22}(1-\beta^2)&0\\0&H_{11}H_{22}(1-\beta^2)\end{pmatrix}, \quad
\mathbf{C}_R^\dag =  
\begin{pmatrix}
H_{11}K_{22} & -H_{22}K_{12} \\
-H_{11}K_{12} & H_{22}K_{11}
\end{pmatrix},
\end{equation}
\begin{equation}
\mathbf{C}_I = \mathbf{0}, \quad
\mathbf{C}_I^\dag = \beta\sqrt{H_{11}H_{22}} 
\begin{pmatrix}
K_{12} & K_{22} \\
-K_{11} & -K_{12}
\end{pmatrix},
\end{equation}
\begin{equation}
\mathbf{C}_R^{(1)} = \mathbf{C}_R - \mathbf{C}_R^\dag \xi_1 = 
\begin{pmatrix}
H_{11}(H_{22}(1-\beta^2)-K_{22}\xi_1) & H_{22}K_{12}\xi_1 \\
H_{11}K_{12}\xi_1 & H_{22}(H_{11}(1-\beta^2)-K_{11}\xi_1)
\end{pmatrix},
\end{equation}
\begin{equation}
\mathbf{C}_R^{(2)} = -\mathbf{C}_R + \mathbf{C}_R^\dag \xi_2 =
\begin{pmatrix}
-H_{11}(H_{22}(1-\beta^2)-K_{22}\xi_2) & -H_{22}K_{12}\xi_2 \\
-H_{11}K_{12}\xi_2 & -H_{22}(H_{11}(1-\beta^2)-K_{11}\xi_2)\end{pmatrix},
\end{equation}
\begin{equation}
\mathbf{C}_I^{(1)} = \mathbf{C}_I - \mathbf{C}_I^\dag \xi_1 = \beta\sqrt{H_{11}H_{22}}\xi_1 
\begin{pmatrix}
-K_{12} & -K_{22} \\
K_{11} & K_{12}
\end{pmatrix},
\end{equation}
\begin{equation}
\mathbf{C}_I^{(2)} = -\mathbf{C}_I + \mathbf{C}_I^\dag \xi_2 = -\beta\sqrt{H_{11}H_{22}}\xi_2 
\begin{pmatrix}
-K_{12} & -K_{22} \\
K_{11} & K_{12}
\end{pmatrix}.
\end{equation}
The Fourier inverse of the matrix $\mathbf{C}(\xi)$ is then
\begin{equation}
\mathcal{F}^{-1}[\mathbf{C}(\xi)] = -\frac{1}{\pi d_2(\xi_2 - \xi_1)} \left\{ \sum_{j = 1}^{2} \mathbf{C}_R^{(j)} T_{\xi_j}(x_1) + \sum_{j = 1}^{2} \mathbf{C}_I^{(j)} S_{\xi_j}(x_1) \right\}.
\end{equation}

\end{document}